# Generative AI: Implications and Applications for Education[1]

*Anastasia Olga (Olnancy) Tzirides, Akash Saini, Gabriela Zapata, Duane Searsmith, Bill Cope, Mary Kalantzis, Vania Castro, Theodora Kourkoulou, John Jones, Rodrigo Abrantes da Silva, Jen Whiting, Nikoleta Polyxeni Kastania*

**Abstract**

*The launch of ChatGPT in November 2022 precipitated a panic among some educators while prompting qualified enthusiasm from others. Under the umbrella term "Generative AI," ChatGPT is an example of a range of technologies for the delivery of computer-generated text, image, and other digitized media. This paper examines the implications for education of one generative AI technology, chatbots responding from large language models (C-LLM). It reports on an application of a C-LLM to AI review and assessment of complex student work. In a concluding discussion, the paper explores the intrinsic limits of generative AI, bound as it is to language corpora and their textual representation through binary notation. Within these limits, we suggest the range of emerging and potential applications of Generative AI in education.*

## 1. Context and History

Chatbots responding from large language models (C-LLM) have sprung to public and educator attention in recent times, notably with the launch of ChatGPT in November 2022. This was version 3.5 of a series of Generative Pre-trained Transformers (GPTs) in development by the company Open AI, founded in 2015. GPT-1was released in 2018, and GPT-4 in March 2023. Microsoft has invested heavily in OpenAI, incorporating its software and data resources into its Bing search, Edge browser and Windows operating system. Google is developing its own GPT, Bard, among many others already developed or with development underway. Images, software code, math and other digitized media can similarly be generated, although these also leverage text for labeling, training and prompts.

    Notwithstanding the hype and anxiety created by these generative AI systems, the underlying technologies are by no means new. The two main components of text-based generative AI are the chatbot that accepts prompts from users and the large language model from which the software draws to frame its dialogical response.

    Chatbots proceed through a human prompt/machine response dialogue. The first computer chatbot was ELIZA, developed in 1964-66 by Joseph Weizenbaum at MIT. To test the program, ELIZA was programmed to ask questions and to respond to answers as if the machine were a psychotherapist along lines developed by Carl Rogers. The art of the psychologist and the intelligence of the chatbot were to reframe information given by the patient as new questions. From a computational point of view, explained Weizenbaum, "[i]nput sentences are analyzed on

the basis of decomposition rules which are triggered by key words appearing in the input text. Responses are generated by reassembly rules associated with selected decomposition rules." However, from the human point of view of its first users, ELIZA uncannily seemed like a psychotherapist. Of the prospects for artificial intelligence, Weizenbaum concluded, "machines are made to behave in wondrous ways, often sufficient to dazzle even the most experienced observer. But once a particular program is unmasked, once its inner workings are explained in language sufficiently plain to induce understanding, its magic crumbles away; it stands revealed as a mere collection of procedures" (Weizenbaum 1966: 36, 43). These generalizations remain as valid today as they were in 1966.

The other principal component of GPTs, the language model, is also a technology that is decades old. The term "language model" is in some ways a misnomer because breakthroughs in this model did not begin until computer-scientists working with language all-but abandoned the project of modeling language, at least in the senses in which humans describe language and account for their usage—semantically (to mean something) and grammatically (its patterning according to differentiated components). Language models "know" nothing about language. They just assess the frequency of character (letter) collocations that may happen to have some semantic and syntactic significance for humans.

However, the emergent phenomena arising from language models of a certain size show that much has in effect been "learned" about language structure and semantic cohesion. Experts in this field are split as to whether GPTs are just behaving as a "collection of procedures" or if there is some mysterious emergent behavior happening that we do not yet understand. Google engineer Blake Lemoine was fired in 2022 for suggesting in an interview with the Washington Post that Google's chatbot initiative, LaMDA, showed signs of sentience (Tiku 2022).

A brief historical recap: When Noam Chomsky began working on his *Logical Structure of Linguistic Theory* in the early 1950s, he posited that language was like a program for human intelligence, in which any particular sentence among the infinite number of possible sentences was "generated by applying optional transformations to the strings underlying kernel sentences" (Chomsky 1956: 123). If language was to be conceived as a program operationalizing human intelligence, it was supposed that a language model could be applied in computing that replicated human reasoning. Versions of this idea are today dismissed as "good old-fashioned AI" (Nilsson 2009: 241).

By the mid 1960s, it became clear that the project of language for application to computing was failing. For the previous decade, the main focus of computational work in linguistics had been machine translation. In 1965, a US government report recommended that this quest should be abandoned because so little progress had been made (Kalantzis and Cope 2020: 209-15).

It was not until the mid 1970s that the first breakthroughs in machine translation were made, now using a completely different approach, statistical text analysis. This, said its early developers, Church and Mercer, in a retrospective overview, is "a pragmatic approach" with an "emphasis on numerical evaluations" focusing on "broad (though possibly superficial) coverage of unrestricted text, rather than deep analysis" (Church and Mercer 1993: 1). Working at IBM in the 1970s, Church and Mercer's boss was pleased with the results of this purely statistical approach. Mercer recalled his boss saying, "Whenever I fire a linguist, the system gets better." A historical aside: After he left IBM, Mercer put his statistical approach to text to powerfully practical effect, first to make a fortune as a hedge fund manager, and later as a driving intellectual force as well as investor in Cambridge Analytica and major funder of the Donald Trump 2016 presidential election campaign (Kalantzis and Cope 2020: 220-21, 235-40).

Since the emergence of GPTs, Chomsky has steadfastly maintained his distance, a paradigm away from statistical approaches to language and artificial intelligence. ChatGPT, he said, is no more than "a lumbering statistical engine for pattern matching, gorging on hundreds of terabytes of data and extrapolating the most likely conversational response or most probable answer" (Chomsky, Roberts and Watumu 2023).

These are the larger dimensions and underpinnings of the debate we analyze in this paper, in theory as well as in educational practice. First, some definitions:

- *Machine intelligence:* following Alan Turing, symbol manipulation by computers larger and more complicated than is feasible for human minds (Turing 1950). Turing demonstrated this in 1949 when the Manchester 1 computer that he and his colleagues had designed found some impossibly large prime numbers (Turing 2015: 212-17).
- *Artificial intelligence:* a term coined by John McCarthy (McCarthy et al. 1955), nowadays taken to mean machines that can learn from human interaction: supervised, unsupervised, reciprocal, and "deep" learning.
- *Generative AI:* coherent and well-formed text, images, and sound generated in new designs by a computer.
- *Chatbot-prompted Large Language Model (*C-LLM)*:* massive textual corpora that can return well-formed textual responses following a human prompt in chatbot dialogue.

We make the case that binary computers in general and C-LLMs in particular are in some respects already enormously "smarter" than humans. Unlike humans they can "remember" huge quantities of digitized texts and images. This makes them potentially very helpful as cognitive prostheses applied in life and learning, providing for instance a narrative response on any topic, an outline of a job application, suggestions about where to eat, and such like. However, there are absolute limits to AI which don't even fall short of human intelligence because they can barely be compared on the same scale. The phrase "Artificial Intelligence" implies that computer intelligence is a replicant of human intelligence. It is not—it is incomparably different, and this is of value. We return to this question in the final section of this paper.

**2. Implications for Education**

By the beginning of the 2020s, there were numerous applications of artificial intelligence in everyday life, from language (e.g., machine translation and voice recognition), to imaging (e.g. face recognition), to social interaction (e.g., the profiles developed in social media), to material practice (e.g. semi-autonomous cars, "smart" home security), and many more. The larger context is a new socio-economic regime sometimes characterized as Industry 4.0, whose signature technologies are artificial intelligence, automation/robotics, bio-informatics, and the internet of things (Schwab 2016). Focusing on the internet specifically, we are entering a phase called Web 3.0, adding artificial intelligence, the semantic web, and blockchain to the social web (Wood 2014). Driven by advanced computing, this is an era of integrated cyber-social systems (Cope and Kalantzis 2022).

However, to date there have been few effective, widespread applications of artificial intelligence to education. The dominant learning management systems, for instance, are still principally file upload/download technologies whose underlying technical and pedagogical architectures have changed little since the beginnings of networked or cloud computing in the 1990s (Cope and Kalantzis 2023b). Some generic applications of AI have been applied in education such as machine translation and grammar and style checking, but these are supplementary external services to support digital text work and not specifically educational technologies.

Nevertheless, education researchers have for some time been exploring the potential impacts of artificial intelligence on education. Lane and colleagues point to the enormous possibilities for what they term AIED in areas such as collaborative, immersive, affective and exploratory learning (Lane et al. 2016). Rosemary Luckin outlines the range of ways in which machine learning can complement human learning (Luckin 2018). A collaborative consisting of eleven leaders in the field have developed an analytical framework to account for the increasing entwinement of artificial intelligence and human learning (Markauskaite et al. 2022). The research team authoring this paper has developed and tested artificial intelligence technologies to support collaborative learning and learning analytics that offer formative feedback (Cope, Kalantzis and Searsmith 2021).

Early research on C-LLMs in education points both to their constructive and worryingly disruptive potentials. Addressing the constructive potentials first: ChatGPT (v.3.5) has been shown to offer "more detailed feedback that fluently and coherently summarizes students' performance than human instructors," demonstrating "high agreement with the instructor when assessing the topic of students' assignments" (Dai et al. 2023). It has been demonstrated that GPTs can be used constructively to support literacy development, from young children's storytelling (Li and Xu 2023) to academic writing (Buruk 2023, Liu et al. 2023). GPT-3 has been shown to perform credibly as the pseudo-teacher in a student-teacher dialogue (Tack and Piech 2022), and presumably this capability will have improved with subsequent releases. GPT-3 was also successfully used to simulate student discourse in dialogue with trainee teachers (Markel et al. 2023), and the significant impact of GPTs in teacher education has been predicted (Trust, Whalen and Mouza 2023).

On the other hand, many educators also fear the disruptive potentials of CLLMs. Open AI's GPT can write a five-paragraph essay on any topic, producing a perfectly well written if dull and predictable response, at least as good or better than a student's response. With more than a hint of irony, a leading AI in Education researcher concluded that GPT-3 will "democratize cheating," putting out of business the expensive essay mills used by an estimated 15% of higher education students who use them (Sharples 2022: 1120). GPT-3.5 was able to pass the US National Board of Medical Examiners test at the level of a third-year medical student (Gilson et al. 2023, Kung et al. 2023) and GPT-4 can pass the Multistate Bar Examination taken by law students (Katz et al. 2023).

This presents an immediate problem for education so long as students have access to GPT—how is it possible to know which parts of student work and thinking where the student's, and which parts were the AI? Asking students to cite the chatbot does not resolve this problem because reliable knowledge claims must be referenced to reliable sources, and there is no knowing which sources a GPT has used. GPT detectors, moreover, are unreliable, and given the generative nature of the AI—every text is a new text—it is unlikely they will ever work reliably. If the cunning student adds a few typos and awkward expressions, they will mostly throw an AI detector off the audit trail.

To "know what students know" (Pellegrino, Chudowsky and Glaser 2001), the immediate solution appears to be rigorously proctored assessment. Some have suggested a return to handwritten submissions, but these could easily be transcribed from an AI generated answer.

"Closed book," proctored assessments are another alternative, but they are less than ideal because they focus on long-term memory, and the validity of this learning measure is today moot. We live and work in knowledge environments where we have come to rely increasingly on externalized memory in the form of the web-connected devices close to our bodies that we constantly need to look up—doctors and lawyers are good examples. Integral to their work nowadays, they do exactly what GPTs do—they need to be able to look things up. We rely increasingly on digital devices as our cognitive prostheses, not only to remember things but to process knowledge in-the-hand with algorithms of calculation and procedure. Today, more than ever before, we need to transfer the center of our focus in education away from long-term memory and towards higher order critical, creative and design thinking that effectively applies long-term social memory to social practice.

Cheating, however, is one of the lesser problems for education created by GPTs. As soon as we expand our notion of knowledge from individual to collective, from personal memory to "cyber-social" knowledge systems (Cope and Kalantzis 2022), we run into much bigger problems with generative AI. On the basis of an analysis of their foundational architecture and algorithmic processes—to the extent that they are disclosed (Open AI 2023)—and on analysis of the development of chatbots and statistical language models, we suggest that C-LLM s are also deeply harmful to a social understanding of knowledge and learning in the following ways.

1. *Sourcing: The machine buries its sources.* Not only are the sources used by C-LLMs opaque, but a user requesting references will be served good-looking but sometimes fake references. In contrast, one of the great intellectual achievements of modern knowledge systems has been to base knowledge claims on the credibility of sources (Grafton 1997). To validate antecedent knowledge claims, we need to be able to interrogate their sources. To distinguish the thinking of the writer from the social knowledge upon which that thinking is based, we use quotation and citation apparatuses. In school, we call this "critical literacy." In academic work, the credibility of sources is dependent on a number of variables, including the qualifications of the researcher, the credibility of the publication venue, and the rigors of peer review. However, giving us the impression that the AI is answering rather than its sources, the sources are hidden. This makes the AI seem smart when it is just a remix of sources. Some of the sources, moreover, may be embarrassing to disclose; others have been copied in breach of copyright (Schaul, Chen and Tiku 2023). The software is a "black box" (Ashby 1956: 86) by design. More recent GPTs have been able to provide some real references when asked, but these are not necessarily the sources from which they have formulated their response.

2. *Facts: The machine can have no notion of empirical truth.* The priority of C-LLMs is to produce convincing narratives. They are genre machines, harvesting ostensible facts they have found in their textual sources but without being able to verify them. They also invent non-existent facts when needed to complete a plausible text (Munn, Magee and Arora 2023). It may happen that some or even most of a response to a prompt is fact, but the machine has no way of distinguishing fact from fake in its sources.

3. *Theory: The machine can have no conception of a theoretical frame or disciplinary practice.* At best, C-LLMs pick up latent semantics in the happenstance of character collocations. They can't know about the connection between dogs and kennels; they just find these character collocations associated under certain textual circumstances. By contrast, disciplinary frames of reference are in human practice rigorously framed ontologies (Cope and Kalantzis 2020: 271-328). These are products of the social intellect, constituted through validated systematic knowledge methodologies that have been codified in scientific and professional practices of observation, multiperspectival corroboration and critical reflection. C-LLMs can do none of these things: they are no more than "stochastic parrots" (Bender et al. 2021), regurgitating text they have copied from a mishmash of textual sources (Magee, Arora and Munn 2022).

4. **Ethics:** *If the machine is socially well mannered, it is not because it sources are necessarily that.* C-LLMs depend on massive textual corpora, and the reality of human legacy text is that the sources are rife with racism, sexism, and homophobia, along with other ideologies and social orientations that are today unacceptable in mainstream public life. To align with the sensitivities and moral agendas of our times and as a necessary corrective to a multitude of existing biases, C-LLMs require extensive filtering. Human programmers create the filters to over-ride the "truth" of the source texts. This is the only way to be sure that the generated texts do not offend modern, liberal sensibilities. However the moral frame of these human-imposed filters is buried. Whether big brother is a nice pseudo-person is less relevant than the fact that C-LLMs are big brothers too, invisible shepherds of our morals.

5. **Critical Dialogue:** *To appear a good interlocutor, the machine is skewed towards being uncritically affirmative.* The "chat" part of the technology of C-LLMs plays through a feigned anthropomorphism. As a good conversationalist, the chatbot remains polite and largely uncritical, even when its human partner is offensive or critical. This, said Weizenbaum, is how "ELIZA maintain[ed] the illusion of understanding." Indeed, "one of its principal objectives [was] the concealment of its lack of understanding" (Weizenbaum 1966: 36). Even the inventor was disturbed by this, a decade later writing a best-selling book renouncing not only chatbots but computer technology in general (Weizenbaum 1976).

If C-LLMs pose a danger for knowledge work and learning in the fundamental areas of their sources, purported facts, theory, explicit ethical frames, and critical dialogue, then how can we apply them in education? In the next section of this paper, we will describe an intervention in which we recalibrate a C-LLM to offer narrative feedback to students.

## 3. Application: The Study

*3.1 Research Design*
In order to explore the potential of C-LLMs in an authentic learning environment, Cope and Kalantzis' Cyber-Social Learning Laboratory[2] at the University of Illinois designed an intervention leveraging a C-LLM to provide learners with AI feedback on extended written texts. This research lab has since 2009 been developing a suite of interconnected apps in a social

---

[2] As well as faculty and doctoral students in Cope and Kalantzis' Learning Design and Leadership program at the university, the Cyber-Social Learning Laboratory hosted visitors including during the period of the intervention described here from the United Kingdom, Greece, Malta, Taiwan, China, and Brazil

knowledge and learning platform, Common Ground Scholar (Cope and Kalantzis 2023a).[3] Research and development work has been built incrementally funded by a series grants from a variety of agencies and foundations.[4]

Our research and software development processes have been closely integrated in a process we have termed "cyber-social research" (Tzirides et al. 2023), bringing together methods of "agile research" inspired by contemporary software development processes (Twidale and Hansen 2019) and design research methods in education (Laurillard 2012). Our masters and doctoral students are as much part of the iterative design process as the research team leaders. During periods of intensive use, new releases of the software were made at midnight almost every day, based on the previous day's interactions with users.

CGMap is a new app within the CGScholar platform, connecting via application programming interface (API) to OpenAI's GPT in order to offer machine feedback to learners on their extended multimodal texts. This machine feedback supplements the peer and instructor feedback provided to students on the same explicitly stated assessment measures. At the point of this trial, CGMap was connected to OpenAI's GPT-3. Since this intervention, it has been connected to later versions. Instructors can enter any assessment into CGMap in the form of a rubric which will provide AI reviews of student work, optionally in addition to anonymized peer-, named instructor- and self-review against the same rubric.

The students participating in our intervention were masters and doctoral students in the Learning Design and Leadership program at the University of Illinois. Sixty-two students joined the trial in two 8-week online courses, "Assessment for Learning" and "New Media and Literacies." The focus of the research was the major project for the course. Students could choose their own topic with a brief to examine educational theory and practice in "a cutting edge area of innovation (such as differentiated instruction, flipped classroom, GPTs, AI in education, learning analytics, gamification, metacognition, self-efficacy/regulation, socio-emotional learning, collaborative learning, formative assessment etc.) or one of education's 'wicked problems' which has presented a longtime challenge (such as a dimension or dimensions of learner diversity, strategies for inclusion, equity and education reform)."

The project workflow is outlined in Fig. 1, with the project commencing in Week 1 of the course, and final works published into the course community and personal portfolios in Week 8.

---

[3] In 2023, CGScholar has nearly 350,000 user accounts and can be integrated into LTI-compliant learning management systems. Parts of the software suite are open to anyone to sign up and use at no charge; other parts have a modest licensing fee based on self-sustainability principles and managed by Common Ground Research Networks, a not-for-profit public benefit corporation based in the Research Park at the University of Illinois. Among others, use cases for CGScholar include: literacy in schools between grades 4 to 12; higher education, including education, engineering, medicine, and veterinary medicine courses; and global social learning interventions by the Red Cross and the World Health Organization.

[4] <u>Learning Analytics:</u> US Department of Education, Institute of Education Sciences: "The Assess-as-You-Go Writing Assistant" (R305A090394); "Assessing Complex Performance" (R305B110008); "u-Learn.net: An Anywhere/Anytime Formative Assessment and Learning Feedback Environment" (ED-IES-10-C-0018); "The Learning Element" (ED-IES-lO-C-0021); and "InfoWriter: A Student Feedback and Formative Assessment Environment" (ED-IES-13-C-0039). Bill and Melinda Gates Foundation: "Scholar Literacy Courseware." National Science Foundation: "Assessing 'Complex Epistemic Performance' in Online Learning Environments" (Award 1629161). <u>Cybersecurity:</u> Utilizing an Academic Hub and Spoke Model to Create a National Network of Cybersecurity Institutes, Department of Homeland Security, contract 70RCSA20FR0000103; Infrastructure for Modern Educational Delivery Technologies: A Study for a Nationwide Law Enforcement Training Infrastructure, Department of Homeland Security, contract 15STCIR00001-05-03; Development of a Robust, Nationally Accessible Cybersecurity Risk Management Curriculum for Technical and Managerial Cybersecurity Professionals, Department of Homeland Security, contract 70SAT21G00000012/70RCSA21FR0000115. <u>Medical Informatics:</u> MedLang: A Semantic Awareness Tool in Support of Medical Case Documentation, Jump ARCHES program, Health Care Engineering Systems Center, College of Engineering, contracts P179, P279, P288.

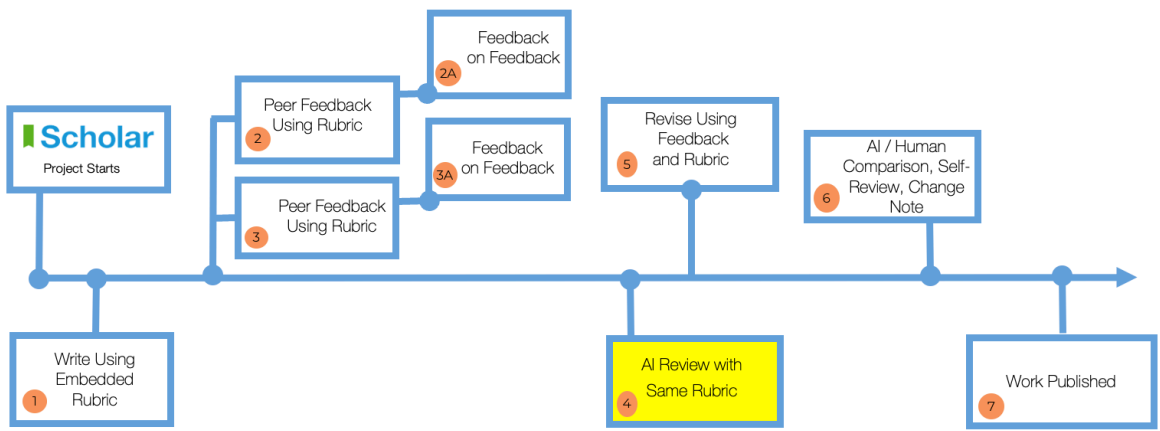

*Fig. 1: CGScholar Project workflow with human and AI reviews.*

An example from one of these works is illustrated in Fig. 2. Course participants commenced writing on the left side of the screen with an elaborated rubric on the right side, against which they write their work. Upon submission of a complete draft, they review two other participant's works against the rubric criteria within the CGMap application. Building on earlier research and development work supporting information and argument writing according to the Common Core Standards in the middle school (Olmanson et al. 2016), the CGMap application consists of a series of review and annotation nodes presented in the form of virtual sticky notes on the right side of the screen, linked to highlighted text on the left. Reviewers connect the nodes into a concept map. The CGMap tool also offers a sentiment analysis feature in reviews and feedback integrated into rubric elements, annotations, general comments, and overall feedback.

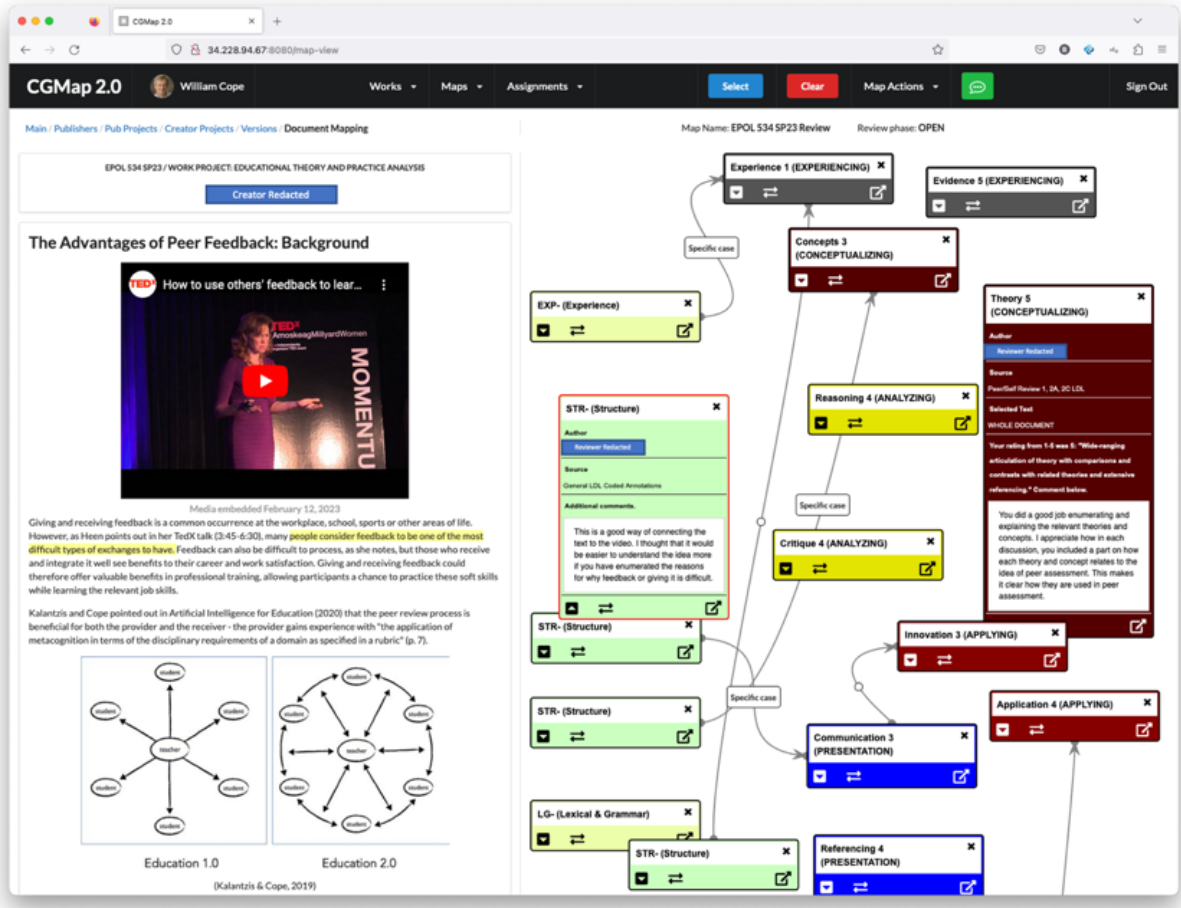

*Fig. 2: One of two peer reviews of version 1 of a course participant's work in CGMap: the work under review is on the left and rubric-based peer review map is on the right. Two nodes are opened in this screenshot, the rest have been folded closed by the reviewer for visual clarity while building the review map. The green node highlighted in red is currently active, an annotation coded as STR- (a structural issue with the text that requires improvement). This refers to the yellow highlighted block of text in the reviewed work. Mousing over the node highlights the text. This annotation coding supports machine learning. The brown node is a whole-text review against the theory review criterion. The reviewer's rating is recorded in the top bar, and their narrative explanation has been opened by the writer.*

After the anonymized peer review, participants offer feedback to reviewers on their reviews, and revise their texts. At this point, a new step had been added to the existing workflow, the AI review. Linked via API to GPT-3, CGMap offered feedback on the same review criteria as the human peers (Fig. 4). Participants then compared the human and machine feedback before final revision and review and publication of their work into the course community and personal portfolios.

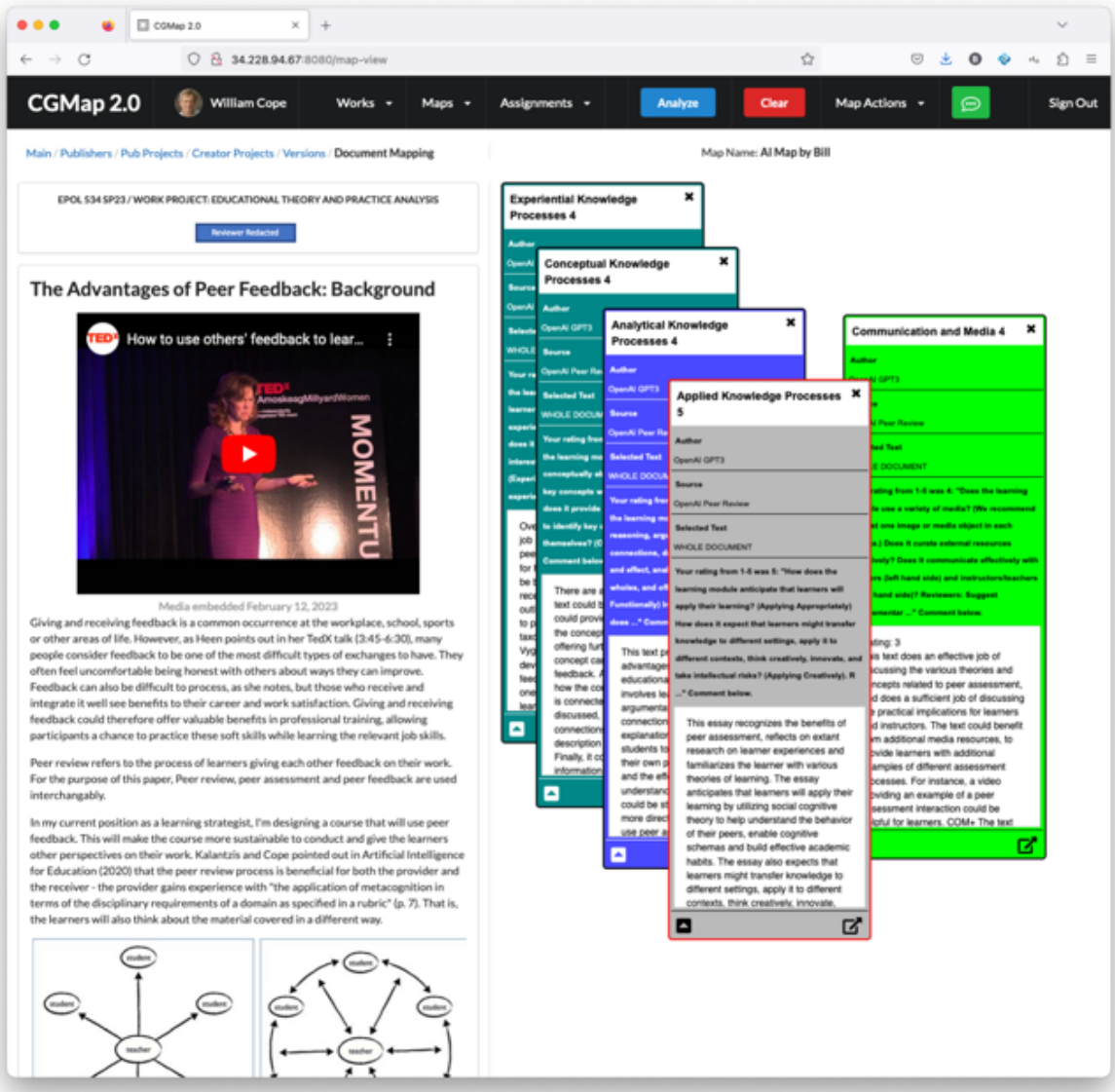

*Fig. 3: AI review of a course participant's work in CGMap*

Here is a brief technical description of the AI review generation process: the CGMap API first takes the extended student text and breaks it into sections. This is helped by CGScholar's strictly structural and semantic markup of sections and subsections, as well as HTML paragraphing at a more granular level. These "sections" are then fed to the ChatGPT-3 "text-curie-001." This model is smaller and less powerful than the model subsequently used for review generation, "text-davinci-003," but it is less expensive, faster, and does a competent job on the summarization task (which would not necessarily benefit from the more powerful model). Each section is summarized, and then sections are concatenated. Summarization was necessary as a result of constraints imposed by GPT-3 but may not be required for future implementations with later versions. After that, criterion by criterion, the summarized text is sent to GPT-3 multiple times, each time with a different rubric criterion as a prompt. The review is displayed as a node graph in the CGMap tool where the AI rubric criterion nodes are generated, one for each criterion on the rubric (Fig. 4). Course participants may then augment the review map with additional information.

The prompt served to GPT consists of the instructions to summarize the student text, system instructions, and a request to review the text based on one of the review criteria in the assessment rubric. This is repeated for however many review criteria there are in the rubric.

For each student work, CGMap can use any rubric provided by an instructor. However in this intervention the rubric items and GPT prompts were drawn from Kalantzis and Cope's "Learning by Design" schema, an overview of which is provided in Fig. 4. Based on the pedagogy of "multiliteracies" (Cope and Kalantzis 2023c, New London Group 1996), this schema takes an epistemological approach to learning, focusing not solely on cognition but more broadly on knowledge-making activities which in addition to cognition involve material practices, embodied activity, and socio-emotional engagement (Cope and Kalantzis 2015, Lim, Cope and Kalantzis 2022). These are high level, abstract review criteria which in theory make the task of the AI more difficult.

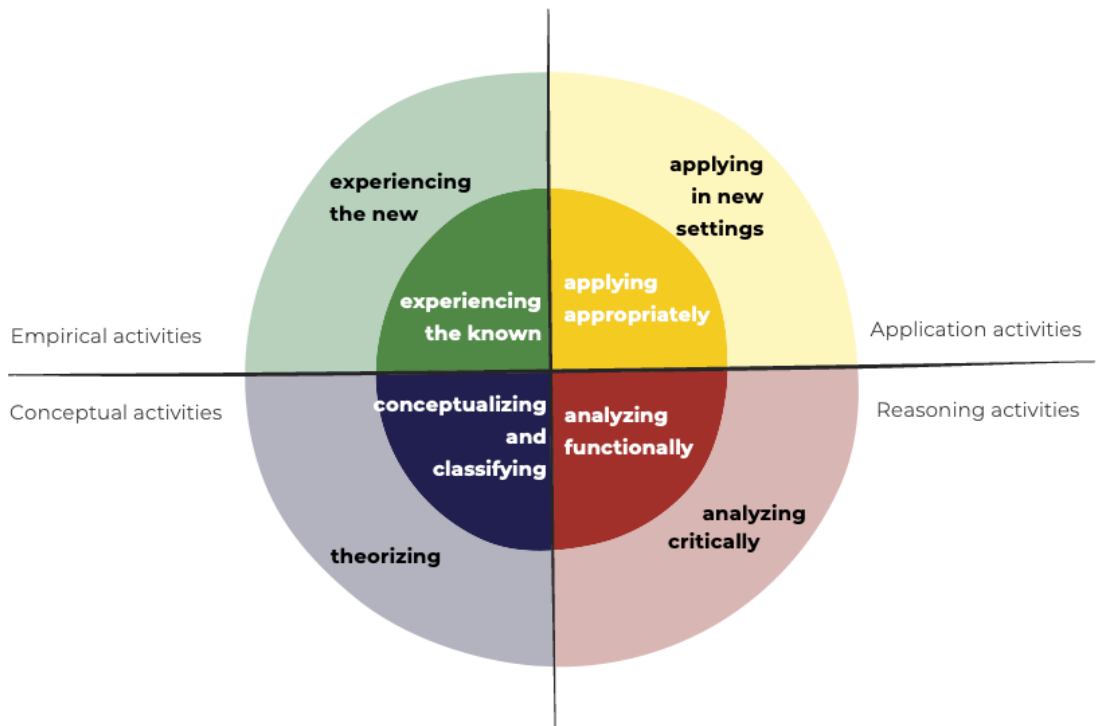

*Fig. 4: The "Learning by Design" Knowledge Processes: A taxonomy of epistemic activities.*

The Learning by Design schema consists of four major areas of knowledge activity, each subdivided into two. *Empirical activities* range from the immediate world, known to the learner (learner experience, positionality, interests, reasons for pursuing a topic), moving into empirical work that immerses learners into new knowledge constituted by systematic observation. *Conceptual activities* involve naming and classifying the world in ways that are often more precisely defined than they are in vernacular language, as well as the disciplinary practice of assembling these concepts into theories. *Analytical activities* apply reasoning to trace functional causes and effects, and critical thinking to explore underlying purposes and interests. *Application activities* test ideas and transfer learning either in predictable ways in typical settings, or creatively into new or unusual settings. For the cognitive aspects of this work, we have created a crosswalk into Bloom's taxonomy, while also highlighting additional socio-material and behavioral aspects that are captured in Learning by Design (Lim, Cope and Kalantzis 2022).

For these eight "knowledge processes" both learners and the AI are provided the same text, the learners in the form of eight rubric criteria the AI as eight engineered prompts. The average length of each rubric criterion is 100 words, including the following: a) a one-sentence definition; b) advice to reviewers on the general kind of feedback that would be most helpful to the author on this criterion; c) marker words typically used to document this particular kind of knowledge activity; and) performance descriptions at five rating levels. When the AI analyzes a learner's text, it runs through the whole text multiple times, analyzing according to each prompt, and returning these to the learner as separate nodes in a concept map, to which they can respond with linked comment nodes.

Course participants worked on their chosen topic for the whole term, making five blog-like updates on the topic on which they were working in the Community app within CGScholar, also commenting there on others' posts. Incremental progress in the course was tracked using CGScholar's analytics app, with each participant having their own progress visualization and the instructors having an aggregated visualization (Fig. 6). At the end of the course, letter grades for the courses were assigned according to each participant reaching announced thresholds in the metrics.

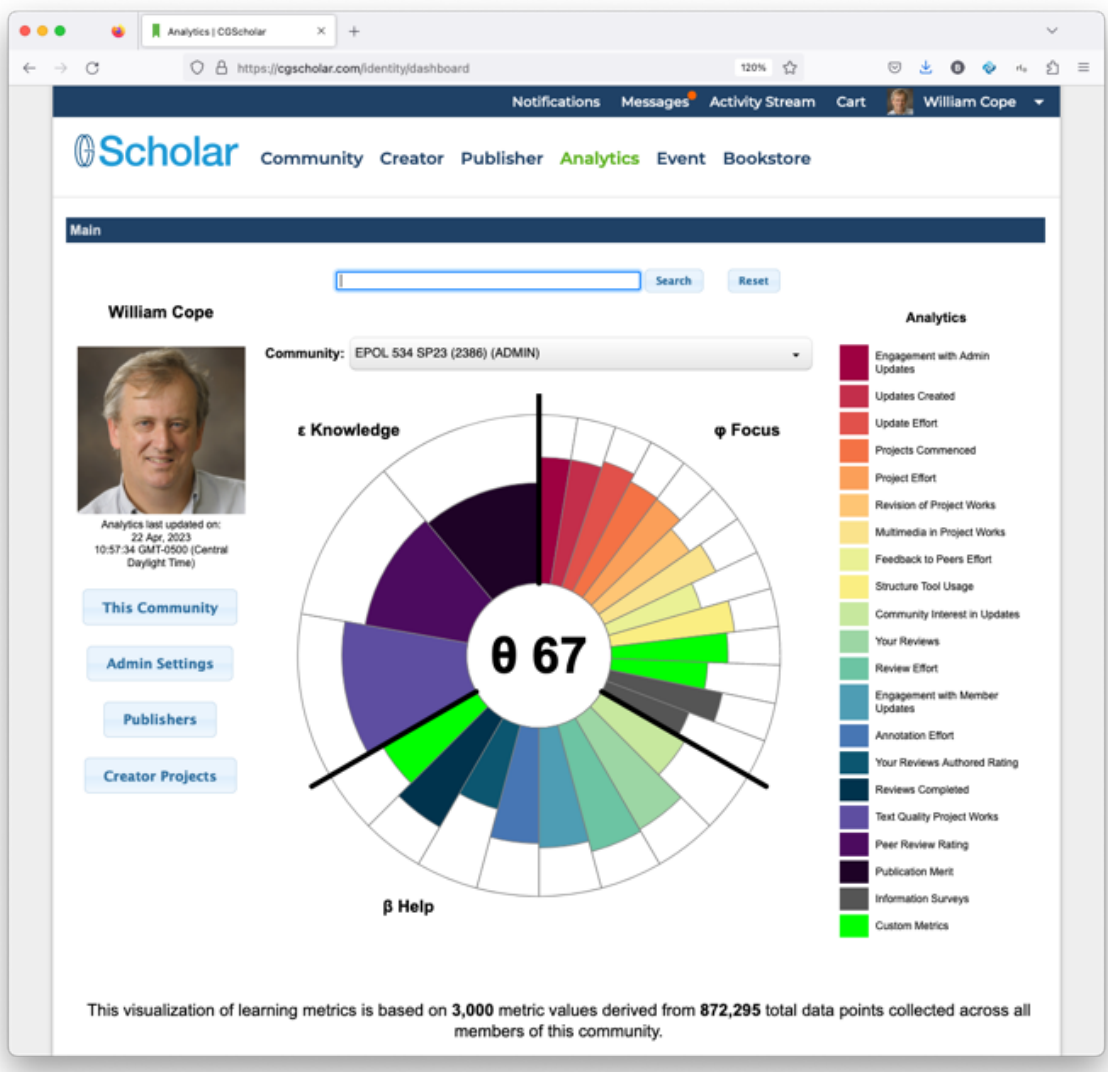

*Fig. 5: CGScholar Analytics App, aggregated instructor view for one of the two courses in this intervention. By the point in the course when this screenshot was taken, the 41 participants had met 67% of course performance expectations in the areas of demonstrable knowledge, collaboration (help), and engagement (focus). The app had mined over 870,000 data points measured across 22 data types and provided 3,000 pieces of actionable machine feedback and machine-mediated peer and instructor feedback. Course participants have the same visualization, just tracking their individual progress.*

3.2 Research Participants
Participants were recruited during Spring 2023 from two online higher education courses in the College of Education at the University of Illinois Urbana-Champaign. The 62 participants were of mixed genders and ethnic/racial backgrounds, ages 25 years and above, pursuing Certificate, Master's, and Doctorate degrees in the College of Education. All were working professional educators undertaking their graduate degrees part-time, highly experienced in areas ranging from elementary school to higher education, workplace and community education, and crossing diverse discipline areas. Recruitment to the research was conducted through announcements in the courses.

*3.3. Research Questions*

The research questions for this project, both for our research team and posed directly to the research participants in this intervention were the following:
1. What are the differences between human and machine reviews?
2. How might these two types of reviews complement each other?

*3.4 Data Collection and Analysis*

The textual corpus that generated the data for this study is summarized in Table 1. Before analyzing these data in the sections that follow, we want to note is that the AI provided more extensive reviews (average length 1335.5 words, or 166.9 words per criterion), than the peers (335.5 words, or 41.9 words per criterion) against the same rubric criterion text.

| Courses | Total Words in Last Version | Avg Words in Last Version | Total Peer Reviews (1-2 per work) | Total Words in Peer Reviews | Avg Words in Peer Reviews | Total AI Reviews | Total Words in AI Reviews | Avg Words in AI Reviews |
|---|---|---|---|---|---|---|---|---|
| New Media and Literacies | 137,505 | 4,436 | 41 | 17,292 | 346 | 29 | 44,592 | 1,393 |
| Assessment for Learning | 172,001 | 4,649 | 54 | 21,436 | 325 | 33 | 42,163 | 1,278 |
| Total or Average | 309,506 | 4,543 | 95 | 38,728 | 336 | 62 | 86,755 | 1,335 |

*Table 1: Extent of the textual corpus: student works, peer reviews, and AI reviews.*

Within and around this textual corpus, the study collected and analyzed the learner experience in four ways, reported upon in the following sections:
a. Comparison of human and AI ratings
b. Sentiment analysis of human and AI ratings
c. Course participants' reactions to the differences between the human and AI reviews
d. Post-course survey evaluating the value and relevance of peer and AI reviews.

*3.4a Comparison of human and AI ratings*

Both human and AI reviews produced ratings using the 1-5 Likert scale. The average ratings obtained from both human and AI reviews for the five rubric elements, as well as the overall average rating, were compared. The text of the reviews was also analyzed for its sophisitication as academic language.

*3.4b Sentiment analysis of human and AI*

To enhance peer feedback and ensure clarity in the intended message, course participants, instructors and the research team had access to sentiment analysis. This is a technology of Natural Language Processing (NLP) that is particularly popular in the field of Educational Data Mining (EDM) (Newman and Joyner 2018, Romero and Ventura 2012). Sentiment analysis helps discern whether a given piece of text has a positive or negative orientation. It involves the use of computational techniques to identify, extract, and quantify affective information from written or spoken language with applications over a wide range of text domains (Wang and Wan 2018). Luo et. al. found in their research that review sentiments expressed in feedback text positively correlated with review scores or rankings, and proposed that sentiment analysis be used as supplementary data to support the validity and transparency of peer review systems (Luo et al. 2022).

CGMap incorporates a feature to connects the app via API to Google Cloud's Sentiment Analysis. Users can call up a sentiment analysis for every node on their map—these are accessed by selecting the double arrows at the bottom of each node, to be seen in Fig. 2. The result is a shadow box with sentiment evaluations, an example of which is provided in Fig. 6.

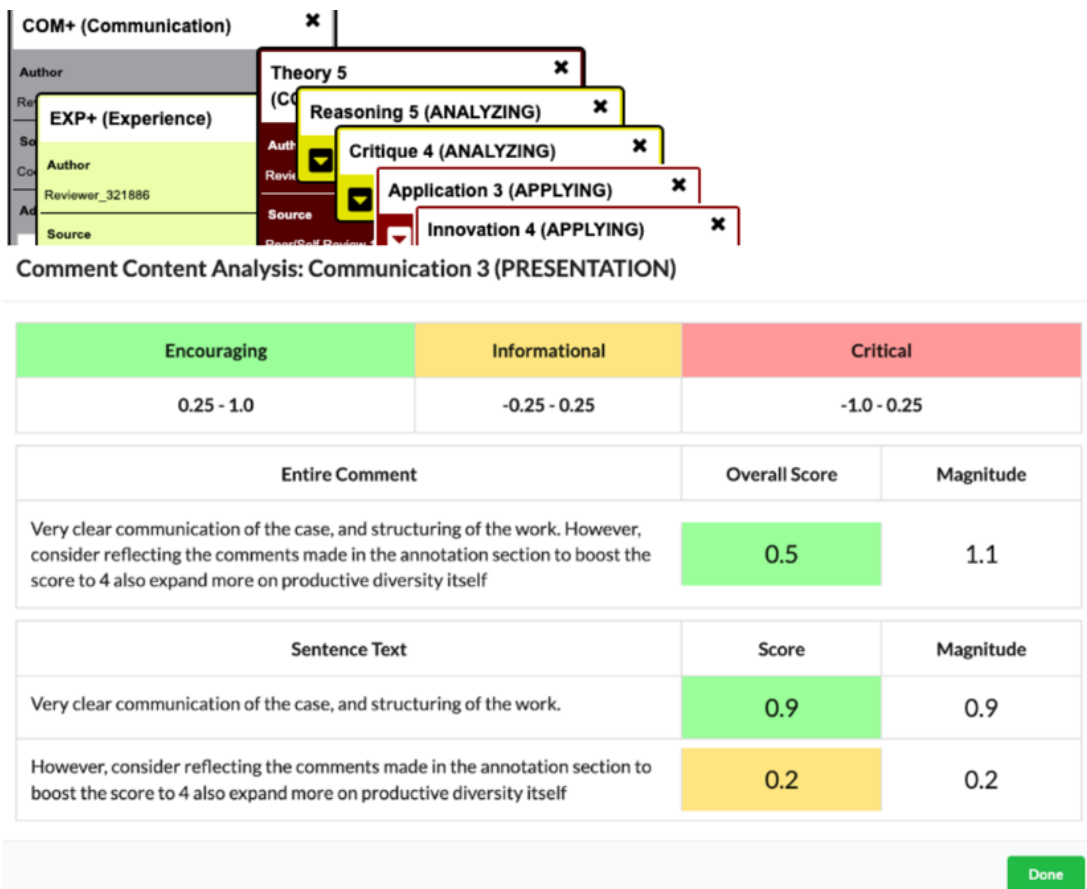

*Fig. 6: Sentiment analysis of a peer's feedback on one of the rubric elements (communication)*

The sentiment analysis determines the overall emotional leaning of the text by providing a numerical score and magnitude values. The score ranges from -1.0 (negative) to 1.0 (positive), indicating the sentiment of the text. The magnitude, on the other hand, indicates the overall strength of emotion, and it is not normalized. It is usually proportional to the length of the text.

The sentiment analysis also provides sentence-level sentiment values, which include score and magnitude values for each sentence. These values help identify the sentiment values expressed in each sentence of the text. In CGMap, the labels "positive", "negative", and "neutral" has been replaced with the color-coded categories of "encouraging", "informational", and "critical" respectively. In addition, the average sentiment score and sentiment magnitude obtained from both human and AI reviews for the five rubric elements, as well as the overall average sentiment score and sentiment magnitude, were compared.

*3.4c: Course participants' reactions to the differences between the human and AI reviews*
After their work had been reviewed by their peers and the AI, participants were invited to express their perceptions of and opinions on their experiences with both review types in the following two ways:
1. For the AI review: Participants were asked to share an image or screenshot of their experience with the AI review and provide three words that summarized their perceptions of this type of review.
2. AI vs. Peer/Human Reviews: Participants were asked to reflect on the main differences between AI and peer/human reviews based on their experience in 150 words or more, and to provide examples illustrated with screenshots.

The students' linguistic responses were analyzed qualitatively with the VERBI software *MAXQDA 2022* using thematic analysis. This type of analysis was chosen because it has been used in a myriad of works that have focused on participants' perceptions and have employed similar instruments for data collection (Braun and Clarke 2006). The first step involved the careful reading of the students' responses and the recording of their overall, general impressions of peer and AI reviews. This was followed by the specific identification of themes and exemplifying statements. In the final stage of the analysis, themes were cross-examined employing Glaser's constant comparative method (Glaser 1965) to ensure that there were no discrepancies in the initial analysis, and the percentage of themes in connection with coded segments for all participants was calculated.

The analysis of students' multimodal responses was grounded in the tenets of social semiotics (Cope and Kalantzis 2020, Kalantzis and Cope 2020, Kress and van Leeuwen 1996 [2021], van Leeuwen 2005). This methodology allowed for the identification and categorization of the semiotic resources (e.g., textual [language and typography], visual, gestural) used by meaning-makers, as well as the examination of their motivation, sociocultural context, and intended audience.

*3.4d Post-course survey*
The post-course survey questions were divided into three sections: demographics, previous experience with peer reviews, and feedback about the peer/human versus the AI reviews. The questions about the peer/human and AI reviews targeted to get students' feedback on the overall quality of both types of reviews, their usefulness in improving the work based on the course rubric, their advantages and disadvantages, the ease of understanding and implementing the reviews' suggestions, and the participants' feelings about having their work reviewed by an AI tool.

*3.5a Results: Human Compared to AI Reviews and Ratings*

In order to prepare the data set for analysis and ensure comparability across variables, we utilized range normalization. This method involved rescaling data by dividing each value in respective rubric element by the range of the entire dataset. The data were reported against five rubric elements, the four macro knowledge processes, "experiential/empirical," "conceptual/theoretical," "analytical/critical," and "applied," plus a measure of academic writing, "communication." We analyzed results against the five rubric elements using the three values namely, 'ratings', 'sentiment score', and 'sentiment magnitude.' Additionally, we employed correlation coefficient and covariance to elucidate the relationships between each element and the overall average for human and AI review respectively.

The resulting bar chart demonstrates that the average rating generated by human reviews was higher than the average rating produced by AI reviews (3.82 vs 3.18 respectively). The variation between the human and AI judgement on each criterion is consistent which seems to indicate that the humans and the AI were in broad agreement about the differences in relative performance (Table 2 and Fig. 7).

|  | Human Review | | | | AI Review | | | |
| --- | --- | --- | --- | --- | --- | --- | --- | --- |
|  | Score | Median (SD) | Correlation | Covariance | Score | Median (SD) | Correlation | Covariance |
| Experiential | 3.54 | 4 (1.44) | 0.71 | 0.85 | 3.00 | 4 (2.12) | 0.39 | 0.92 |
| Conceptual | 4.01 | 4 (0.96) | 0.81 | 0.65 | 3.13 | 4 (2.29) | 0.63 | 1.62 |
| Analytical | 3.77 | 4 (0.92) | 0.82 | 0.63 | 3.10 | 4 (2.16) | 0.56 | 1.36 |
| Applied | 3.83 | 4 (1.00) | 0.86 | 0.72 | 3.19 | 4 (1.99) | 0.61 | 1.36 |
| Communication | 3.97 | 4 (1.03) | 0.73 | 0.62 | 3.48 | 4 (1.77) | 0.51 | 1.01 |
| Average | 3.82 | - | - | - | 3.18 | - | - | - |

*Table 2: Human and AI ratings on review criteria*

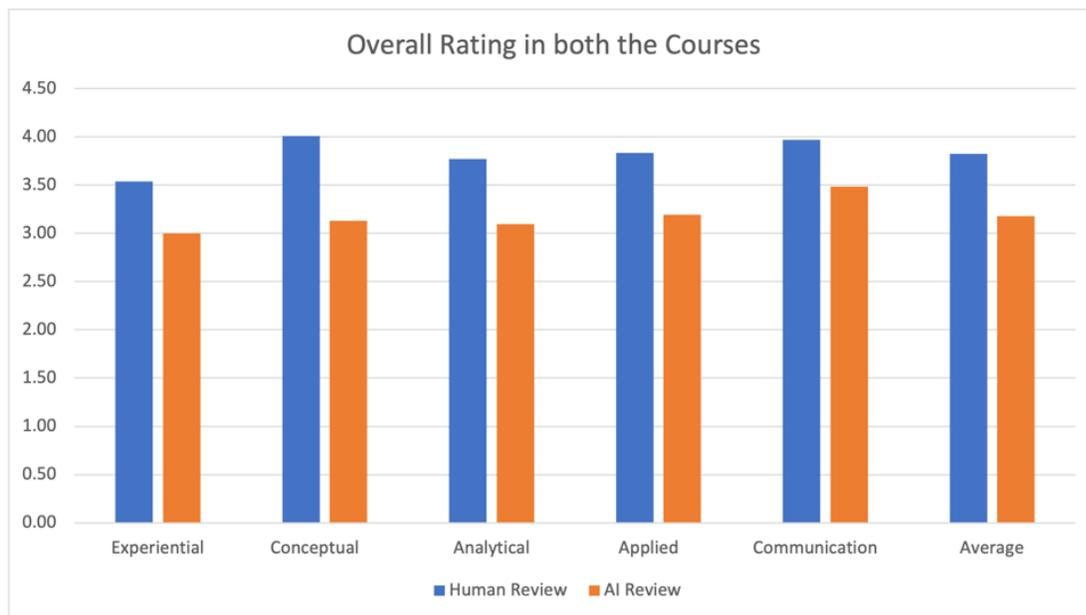

*Fig. 7: Human and AI ratings on 1-5 Likert scale*

Our findings also indicate that human peer review generated slightly higher ratings than AI review against identical rubric elements and rating level descriptions. In other words, the AI reviews were somewhat more critical and the human reviews somewhat more favorable. How does one interpret this result given that chatbots are designed to be more agreeable than argumentative? Among possible explanations, while the prompts specifically directed the chatbot to provide critical commentary, the human reviewers tended to be generous to each other given that they belonged to the same community of learners and even though the reviews were anonymous.

Further, we observed that the median values (Fig. 8) for the ratings variable were skewed to the left. In terms of variability within the data, we found that the standard deviation (SD) for human peer reviews was generally smaller than that of AI reviews, while the data in AI reviews were more spread out. The smaller SD values in human peer review ratings may indicate a more consistent review method, while the larger SD values in AI review dataset may imply greater variability in the results obtained using the other review method. This may also indicate that the "window" of possible scores, in this context, is skewed towards higher values thus compacting the score range.

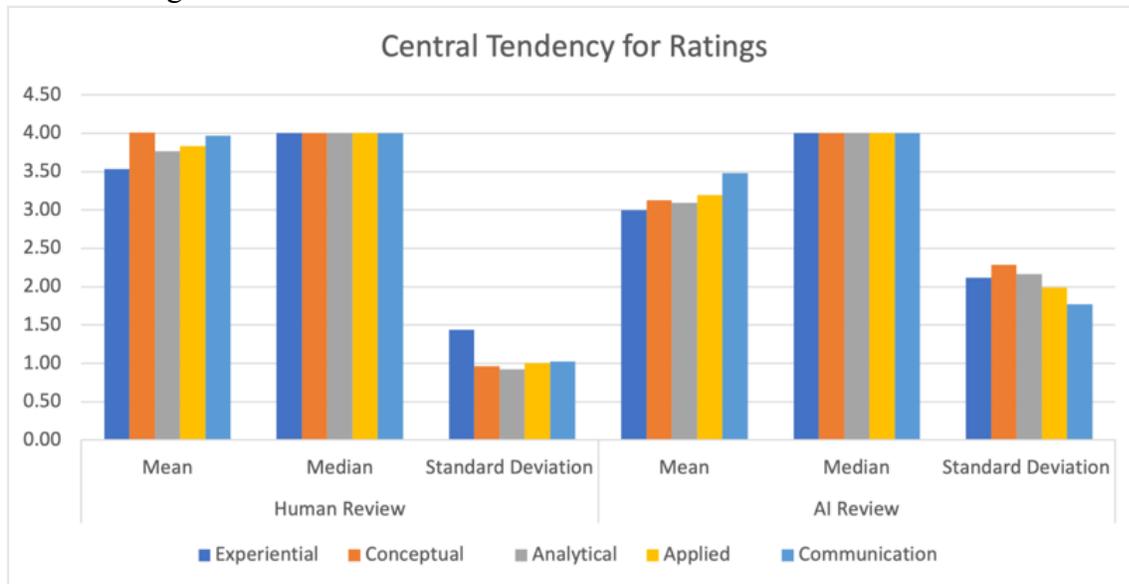

*Fig. 8: Median and Variability of human and AI ratings*

Additionally, when comparing the covariance (Fig. 9) which highlights the direction of the linear relationship in datasets of human peer review and AI review, we can see that the values for human peer review seem to be relatively small, suggesting a weak or even non-existent uni-directional linear relationship. On the other hand, the values for AI review are generally larger and positive, indicating a stronger linear relationship. This suggests that the ratings in human peer review may not be strongly varying together, while the ratings in AI review more strongly vary together.

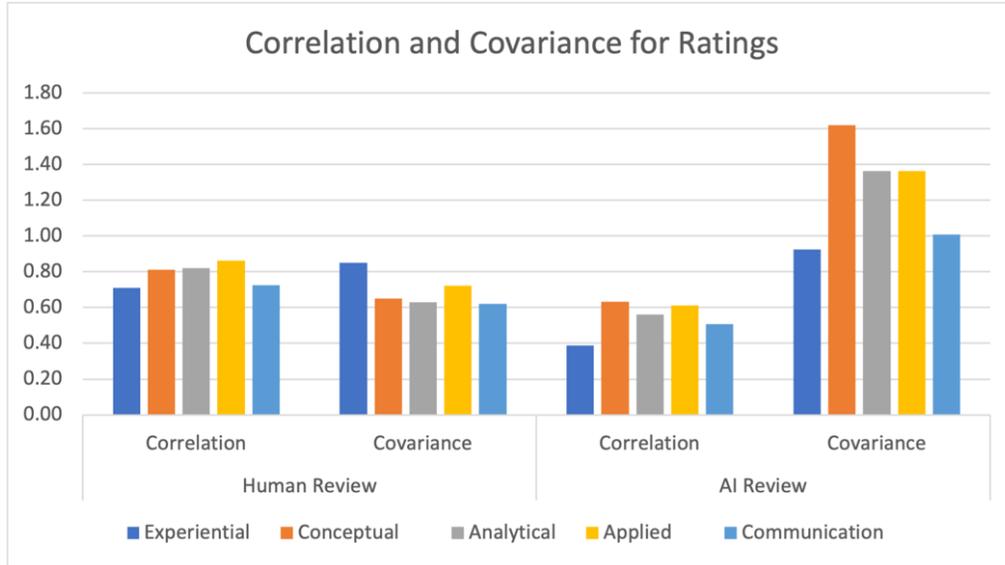

*Fig. 9: Correlation and Covariance between the rubric elements and their respective average*

Correlation coefficients (Fig. 9) showcase intensity among the elements in the datasets of human peer review and AI review. The values in human peer review dataset are consistently higher than the values in AI review dataset for both the courses. This suggests that the five rubric elements being studied in dataset human peer review are more strongly correlated than those in AI review dataset which may therefore have a stronger impact on student success than the factors in the AI review dataset. The correlation values in the AI review data set fluctuate more than those in the human peer review, with some having relatively low correlations and others having relatively high correlations. This could indicate that the relationship between elements in AI review are more complex or that there may be some outliers or influential data points affecting the observed correlations.

Assessing the quality of the AI compared to the human text, we used CGScholar's Analytics app's measure of academic language. This combines readability measures: Flesch–Kincaid, Coleman–Liau and the Automated Readability Index. The number approximates a school grade level. Together, these three measures capture the complexity and level of sophistication of syntax and vocabulary. We take these to be a proxy for the academic language which is typically more complex on these measures. A comparison of the full corpus of human and AI reviews (Table 3) reveals complex language use in the AI reviews than the human reviews, with higher mean (15.80 vs 7.80), median (16.25 vs 7.95), and maximum scores (18.03 vs 15.42) for AI reviews, The difference in standard deviation between the two groups is small, suggesting that there is not a significant difference in the consistency of readability levels.

|  | Human Review | AI Review |
| --- | --- | --- |
| Mean | 7.80 | 15.80 |
| Median | 7.95 | 16.25 |
| Maximum | 15.42 | 18.03 |
| Standard Deviation | 2.71 | 2.43 |

*Table 3: Human and AI reviews: readability level measures as a proxy for complexity in academic language*

*3.5b Results: Sentiment Comparison, Human and AI Reviews*

Sentiment analysis of the text of the human and AI reviews are presented in Tables 4 and 5. This comprises sentiment score (between -1 and +1), and sentiment magnitude (between 0 and +inf):

|  | Human Review | | | | AI Review | | | |
| --- | --- | --- | --- | --- | --- | --- | --- | --- |
|  | Sentiment Score | Median (SD) | Correlation | Covariance | Sentiment Score | Median (SD) | Correlation | Covariance |
| Experiential | 0.24 | 0.30 (0.42) | 0.51 | 0.05 | 0.19 | 0.20 (0.21) | 0.72 | 0.02 |
| Conceptual | 0.43 | 0.40 (0.39) | 0.74 | 0.07 | 0.28 | 0.30 (0.16) | 0.66 | 0.01 |
| Analytical | 0.25 | 0.25 (0.31) | 0.64 | 0.05 | 0.23 | 0.20 (0.20) | 0.78 | 0.02 |
| Applied | 0.33 | 0.35 (0.33) | 0.70 | 0.06 | 0.20 | 0.20 (0.16) | 0.71 | 0.01 |
| Communication | 0.40 | 0.40 (0.39) | 0.67 | 0.06 | 0.20 | 0.20 (0.16) | 0.38 | 0.01 |
| Average | 0.33 | - | - | - | 0.22 | - | - | - |

*Table 4: Sentiment Score on review criteria*

|  | Human Review | | | | AI Review | | | |
| --- | --- | --- | --- | --- | --- | --- | --- | --- |
|  | Sentiment Magnitude | Median (SD) | Correlation | Covariance | Sentiment Magnitude | Median (SD) | Correlation | Covariance |
| Experiential | 0.89 | 0.80 (0.59) | 0.62 | 0.16 | 3.83 | 3.65 (1.93) | 0.52 | 0.87 |
| Conceptual | 1.09 | 0.90 (0.59) | 0.64 | 0.16 | 3.48 | 3.20 (1.40) | 0.55 | 0.67 |
| Analytical | 0.93 | 0.85 (0.50) | 0.67 | 0.14 | 2.57 | 2.30 (1.60) | 0.62 | 0.85 |
| Applied | 0.90 | 0.75 (0.51) | 0.66 | 0.14 | 2.62 | 2.30 (1.53) | 0.67 | 0.87 |
| Communication | 1.58 | 1.4 (0.94) | 0.76 | 0.30 | 2.45 | 2.2 (1.28) | 0.39 | 0.43 |
| Average | 1.08 | - | - | - | 2.99 | - | - | - |

*Table 5: Sentiment Magnitude on review criteria*

Our analysis of the human and AI reviews (Tables 4 and 5) revealed that the average sentiment score from human reviews was slightly higher than that of AI reviews (0.32 vs 0.22), but AI review consistently outperformed human peer review in terms of sentiment magnitude for all rubric elements (see Table 5) which is a measure of the strength or intensity of the sentiment expressed in the feedback (Fig. 10). The AI review generated significantly higher sentiment magnitude compared to human peer reviews (1.08 vs 2.99), possibly due to its ability to provide longer and more in-depth feedback that consistently followed the prescribed prompt. These findings suggest that AI review may generate more detailed and informative feedback to students compared to human reviews.

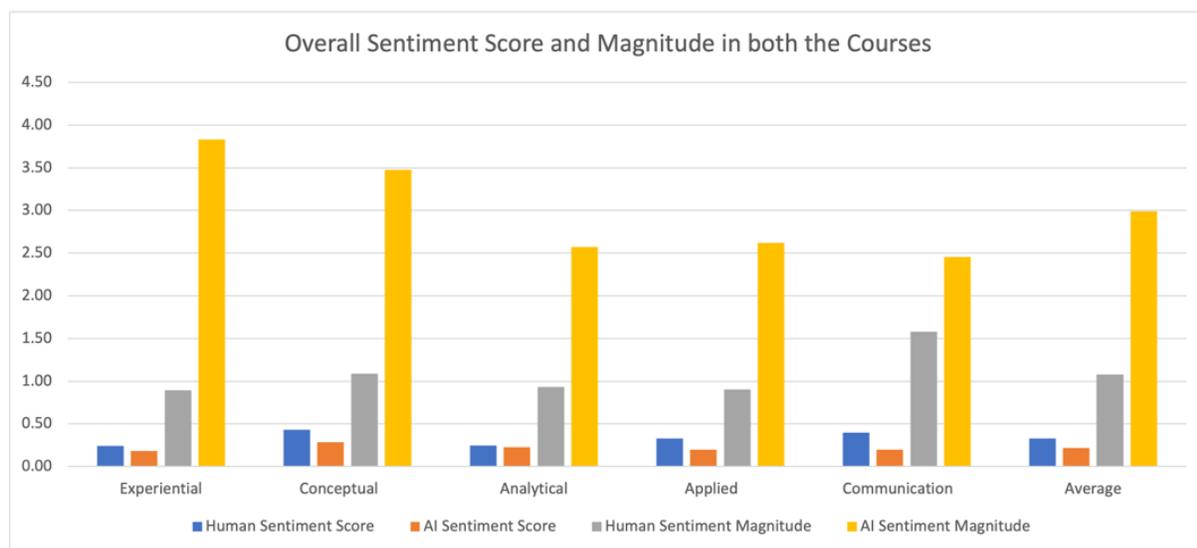

*Fig. 10: Sentiment score and magnitude for both courses*

We also observed that, for the sentiment score (Fig. 11) and magnitude (Fig 12), there was no clear skewing of median values. The data seems to be relatively symmetrical in some sets, while in other sets, it is skewed to the right or left as indicated the charts below. In terms of variability within the data, we found that the SD values were again in flux—low for sentiment score (Fig. 11) indicating a narrow range of values and high for sentiment magnitude (Fig. 12) indicating a wide range of values. These findings suggest that the AI review could potentially be a more objective measure of student performance but may not be as effective as human reviews given the larger differences in mean and median scores and the relatively higher standard deviation for AI reviews indicating a greater variability in the results.

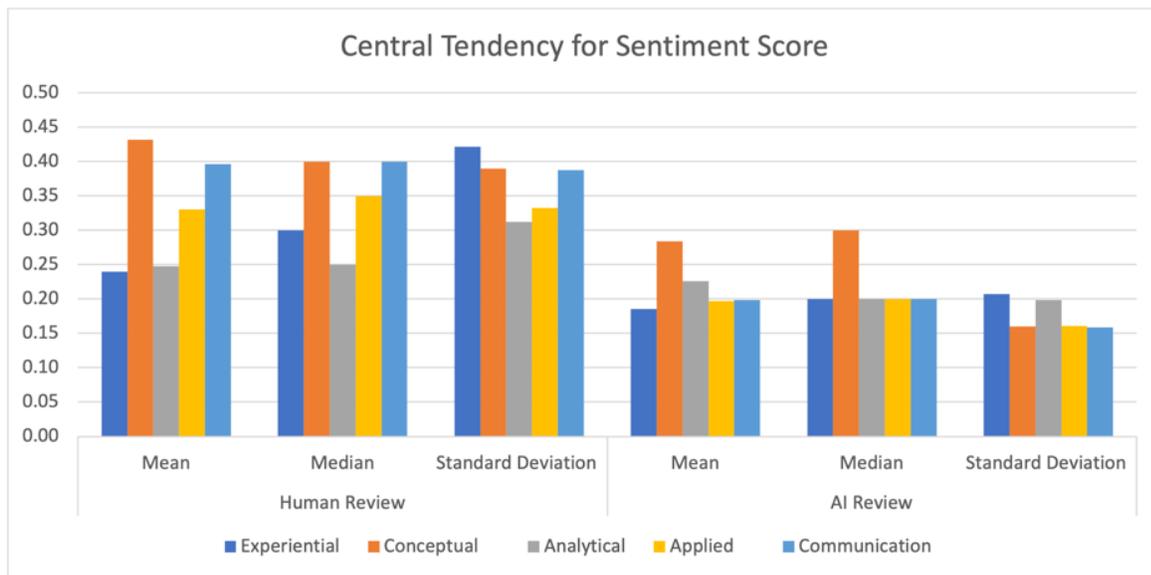

*Fig. 11: Sentiment scores, human and AI reviews*

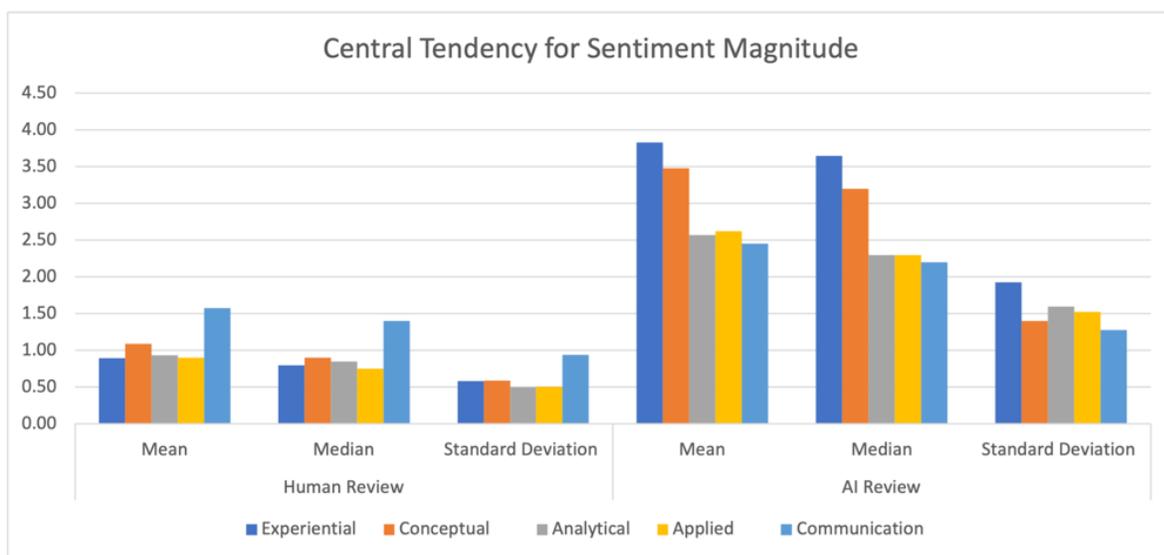

*Fig 12: Sentiment magnitude, human and AI reviews*

When comparing the covariance of the sentiment score (Fig. 13), we observed that the values for AI review seem to be relatively small, suggesting a weak or even non-existent unidirectional linear relationship between the variables. The values for human peer review are generally larger and more positive, indicating a stronger linear relationship. This suggests that the sentiment score in AI review may not be strongly varying together, while in human peer review, it varies more strongly. Conversely, for the covariance of the sentiment magnitude (Fig. 14), we can see that the values for human peer review seem to be relatively small, suggesting a weak or even non-existent unidirectional linear relationship between the variables. The values for AI review are generally larger and more positive, indicating a stronger linear relationship. This suggests that the sentiment magnitude in human peer review may not be strongly varying together, while in AI review, it varies more strongly which could be influenced by the greater average length (Table 1).

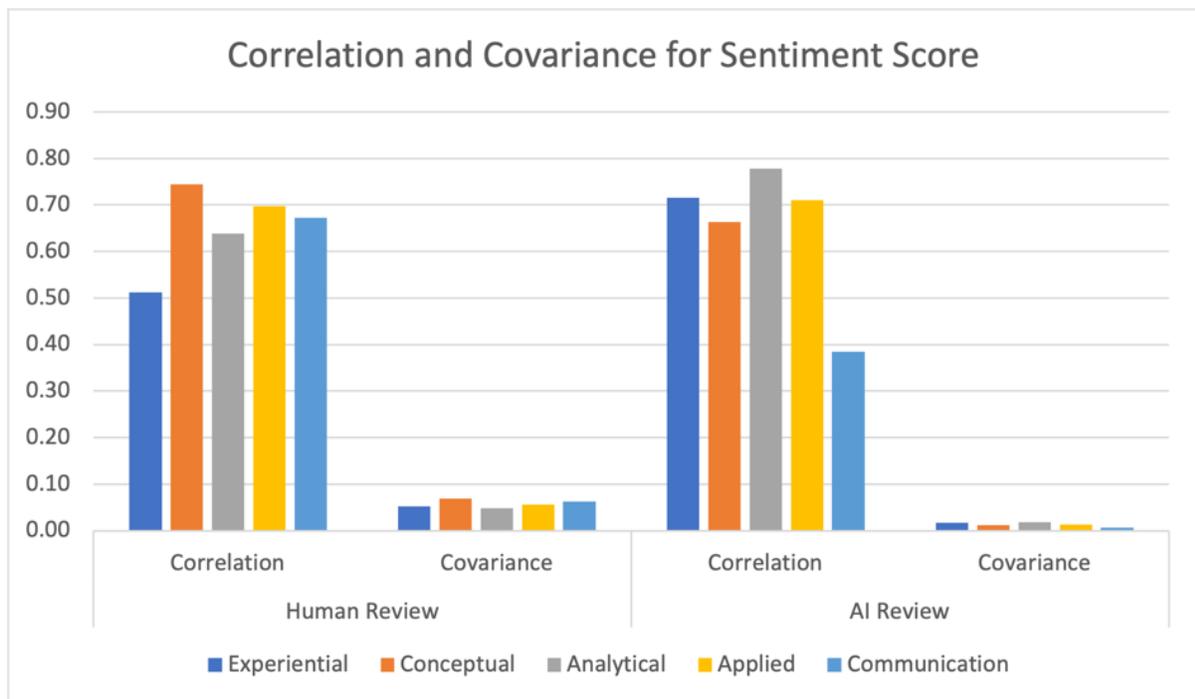

*Fig. 13: Correlation and Covariance of sentiment analysis and score*

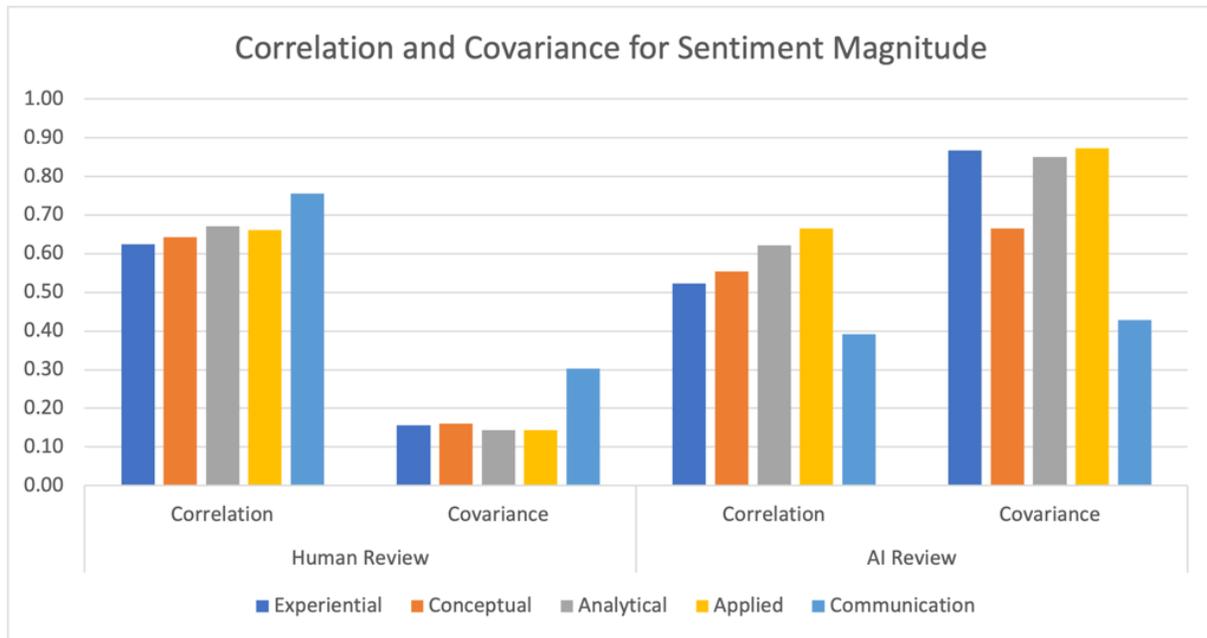

*Fig. 14: Correlation and Covariance of sentiment sore and sentiment magnitude.*

When comparing the correlation coefficients of the sentiment score and sentiment magnitude respectively, we found that the values in the human peer review dataset are relatively similar to the values in dataset AI review (Fig. 13 and 14). Just like in the ratings, the correlation values of sentiment score and sentiment magnitude in the AI review dataset are more fluctuating than those in the human peer review, with some having relatively low correlations and others having relatively high correlations. This could indicate that the relationship between elements in AI review are more complex or that there may be some outliers or influential data points affecting the correlations.

In summary for the rating, review and sentiment analyses, we found that the average rating generated by human reviews was higher than that produced by AI reviews, with human reviews having smaller standard deviation, indicating a more consistent review method. Our findings also suggest that human reviews generated a slightly higher average sentiment score than AI reviews, but AI reviews consistently outperformed human reviews in sentiment magnitude, indicating the ability to provide longer and more in-depth feedback. The standard deviation of sentiment score and sentiment magnitude suggests that AI reviews with a wider range of values could be a more objective measure of student performance, but not as effective as human reviews. The covariance values for ratings, and sentiment magnitude suggest a stronger linear relationship between the variables in AI reviews, while human peer reviews show a weaker or even non-existent relationship. In the case of sentiment score, AI review reviews show a weaker or even non-existent relationship which suggests that the sentiment score in AI review may not be strongly varying together, while in human peer review, it varies more strongly. The correlation values for ratings, sentiment score, and sentiment magnitude in AI reviews are more fluctuating than those in human peer reviews, indicating a more complex relationship or that there may be some outliers or influential data points affecting the correlations.

Overall, the analysis suggests that human peer reviews outperform AI reviews in terms of accuracy and consistency in the ratings category. However, the results from sentiment score and magnitude suggest that AI reviews could be a valuable tool for providing detailed and informative feedback to students. Therefore, a combination of both may be useful in providing a more comprehensive evaluation of student performance.

*3.5c Results: Course Participants' Perspectives on the AI Reviews*
The results of the thematic analysis of the participants' textual responses revealed themes connected to both benefits and drawbacks in peer and AI reviews. A summary of the themes resulting from the qualitative analysis and their percentage with respect to the coded segments in all responses is presented in Fig. 15.

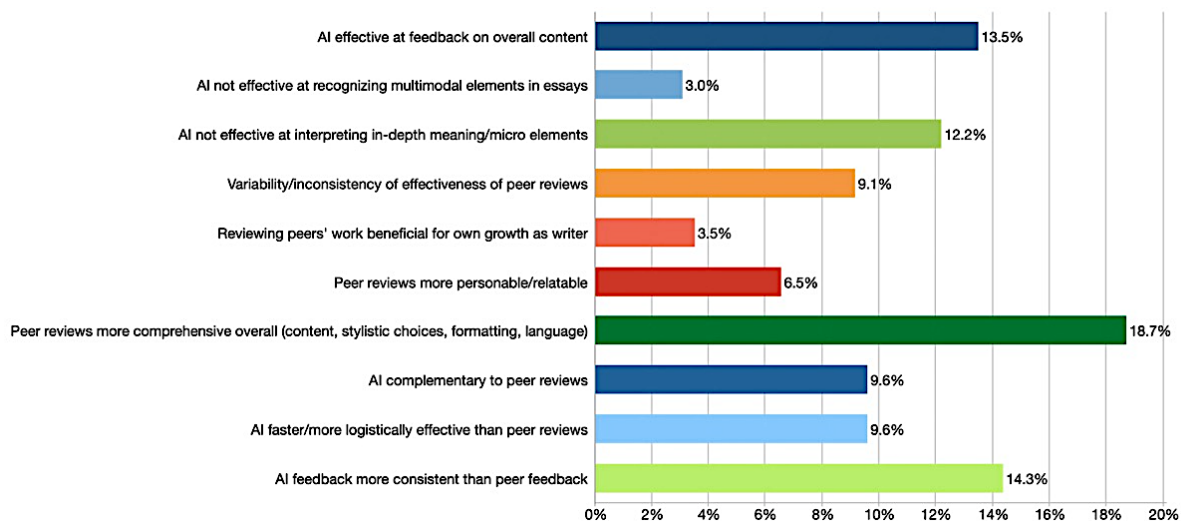

*Figure 15: Frequency of themes in coded segments*

Most course participants believed peer reviews were more beneficial than their AI counterparts, with 7 of them stating that they would prefer only human feedback. These opinions were based on the comprehensive and in-depth feedback resulting from peer feedback with respect to essay content, formatting, and stylistic choices:

"[The peer review] is not only specific but demonstrates that the reviewer has the entirety of the work and its purpose in mind. I think this is a key distinction between human peer reviews and their current AI counterparts: a human reviewer has a thesis, a guiding idea behind their critique that they form over the course of reviewing the work, while an AI does not" (Participant 4).
"During [the] human review, I noticed that the feedback displayed a level of understanding and sensitivity to [the] issue [discussed]. It felt like the person providing feedback had a more holistic view of my work and evaluated it accordingly" (Participant 19).

Another benefit noted was the growth participants had experienced in their own writing as a result of reviewing their peers' work: "Giving a peer-review is time consuming on the reviewer's part, but I also see it [as] being a productive part of the process. I have learned so much reviewing others' work. For starters, reading the broad topics have made me a better learner. I am now a more professional writer" (Participant 13).
These themes were also reflected in the words chosen by the participants to summarize their experiences with peer reviews. Some of the words had a clear positive connotation (e.g.,

"supportive," "meaningful," "thoughtful," "encouraging," and "helpful") and seemed to focus on the human, emotional aspects of the reviews. Other words had a more negative connotation (e.g., "disappointing," "spotty," and "lacking") and appear to signal the lack of quality and time delay in the comments received by some course participants. Some terms also emphasized the comprehensiveness and usefulness of human reviews (e.g., "thorough," "specific," "constructive," "targeted," "enlightening," and "purposeful"). Overall, the tone of the words chosen was more positive than negative, confirming the participants' preference for this type of review.

When considering the effectiveness of the AI reviews, the students in this work noted that the comments received had only been useful for identifying "big picture" revisions based on the rubric categories, but they had been limited in terms of meaningful feedback due to the AI's lack of understanding of the context and nuances of academic writing:

> "I thought the feedback was really vague and too generalized… It could not tell me where I needed to add more details, better structure, or deeper analysis" (Participant 8).
> "The feedback I received on my work seemed general and could [have] been applied to any number of works. It did a great job summarizing, but not interpreting, my writing" (Participant 12).

Despite these drawbacks, the AI reviews were praised for their speed and overall efficiency as well as for the consistency of the feedback provided, which contrasted with the variability found in the quality of peer comments and their lack of expediency:

> "I could not believe how quickly the review was created. When I read through the review, I was pleased to find that all the comments were rubric aligned and pointed out… ways to continue to improve" (Participant 30).
> "AI reviews are… more consistent than human reviews" (Participant 36).

Because of these advantages, five of the participants stated they would choose these types of reviews over peer feedback. Nevertheless, since peer and AI reviews appear to offer different kinds of benefits, most participants thought they could complement each other, instead of one type of feedback being replaced by the other:

> "The combination of both a peer and AI review was a nice mixture and balance that provided pragmatic feedback that was used when reassessing my work" (Participant 7).
> "I feel like human review should not be replaced with AI review but they can work alongside each other" (Participant 26).

The words used by the participants to describe the AI feedback seem to focus on its objective and impersonal, data-driven nature as well as its convenience: "fast," "straight-forward," "immediate," "practical," "satisfying," and "instant." These data suggest that the AI comments were easy to understand, and they provided clear and actionable information or insights as well as instilled a sense of progress or momentum in participants with respect to the development of their work. On the other hands, words like "superficial," "decontextualized," "general," "rough," and "disjointed" convey a sense of insufficiency and lack of substance or detail, as well as a lack of focus, sophistication, or coherence in the reviews. In spite of these weaknesses, most participants appear to welcome and enjoy this type of reviews, as long as they did not replace peer reviews, but complemented them.

Like the textual responses, the 29 multimodal elements submitted by the participants to describe their experiences with peer and AI reviews highlighted advantages and drawbacks. Additionally, some of the artifacts conveyed students' uneasiness, anxiety, and ambivalent emotions towards the use of AI for educational purposes, while others expressed the opposite, celebrating instead the potential for human-AI collaboration. The semiotic resources in all products were combined in single, non-segregated frames (i.e., all the semiotic resources were presented together in one semiotic space). The prevailing modes of communication were the visual, spatial, and gestural (though a limited number included some textual features).

The main advantage of peer reviews highlighted by the participants' multimodal artifacts was their diverse and collaborative nature. This was achieved through the use of different colors, photos or illustrations of diverse students, spatial and gestural features denoting proximity and equality, and facial expressions and body language conveying dialogue, harmony, and a friendly exchange of ideas. While the diversity of peer reviews appears to have been welcomed by some course participants others conveyed more negative emotions, as the participants who submitted them seem to have found the diversity of opinions negatively subjective and overwhelming (i.e., it was difficult to reconcile comments that differed quite drastically) and the reviewers' unreliability and/or delay in terms of the submission of their feedback. One image depicted an arm and a hand holding a pen that act as a visual and gestural synecdoche for the writer, who appears to be buried in paper reviews. The writer, nevertheless, is shown as breaking through the reviews, emerging triumphally. This is conveyed gesturally by the fist gripping the pen in a gesture symbolizing triumph,

Of the multimodal artefacts submitted by the participants to convey their experiences with non-human reviews, half presented the AI tool as isolated robots with humanoid features, some located in the metaverse (presented in several images as an abstract, dark, and/or vacuous space). These robots exhibit neutral facial expressions, and are either looking at a particular point within the setting where they are or are staring blankly into the metaverse: There is no visual contact between the figures and viewer. These visual and gestural elements appear to highlight the impersonal, decontextualized "cold" tone of the AI reviews, which contrasts with the collaboration and communication emphasized in the artifacts depicting peer reviews.

Other course participants presented images that portrayed the AI as complementary to peer feedback, submitting multimodal artifacts that emphasized AI-human communication and collaboration. These artifacts included photos or illustrations of robots shaking hands with humans or making physical contact with them (e.g., fingers touching). The representations relied on visual and gestural synecdoche, as only one arm or an arm and a hand for each figure is shown (i.e., no full-bodied robots or human beings are included in the images chosen), which could be interpreted as a generalized view of this relationship. Also important is the fact that in all images, the technology and human elements are the same size, and though they might originate from opposite points (e.g., left vs. right), they converge on a central point. This could mean that, for these students, AI and peer reviews could complement each other and could equally contribute to the improvement of their work.

Some participants expressed their anxiety and ambivalent feelings towards the use of AI in educational contexts. To convey their views, they resorted to images from movies that were either created with live-action and motion-capture computer-animated animation or featured a villainous computer. For example, Participant 4 used an image from the film *The Polar Express*, whose type of animation has been associated with Mori's concept of the uncanny valley (Russell 2021). Masahiro Mori was a leading Japanese professor of robotics who posited that "a person's response to a humanlike robot would abruptly shift from empathy to revulsion as it approached, but failed to attain, a lifelike appearance" (Mori 1970 [2012]: 98). This resonates with Sigmund Freud's famous essay on the uncanny, capturing a moment when doubts are raised in a person's mind "whether an apparently animate being is really alive; or conversely, whether a lifeless object might not be in fact animate" (Freud 1919 [1955]: 226). By choosing an image from a movie that has been connected with this concept, the participant drew parallels between the uneasiness produced by the uncanny valley and their experience with AI.

*3.5d Results: Post-Course Survey*
The survey data indicated that most students (n=24) felt curious about having their work reviewed by an AI tool, while some (n=14) were excited about using it and only a few students (n=3) stated that were indifferent about it.

Considering the overall quality and the usefulness rating of the reviews, the human reviews were rated slightly higher in both domains compared to the AI reviews, as Fig. 16 and Fig. 17 show. Regarding the ease of comprehension and implementation of the review suggestions in the revision of their works, the majority of students (n=30 for the AI reviews and n=32 for the human reviews) rated the process as 'very' or 'somewhat' easy, which showcases that there is no difficulty in deploying either type of feedback to the learning process.

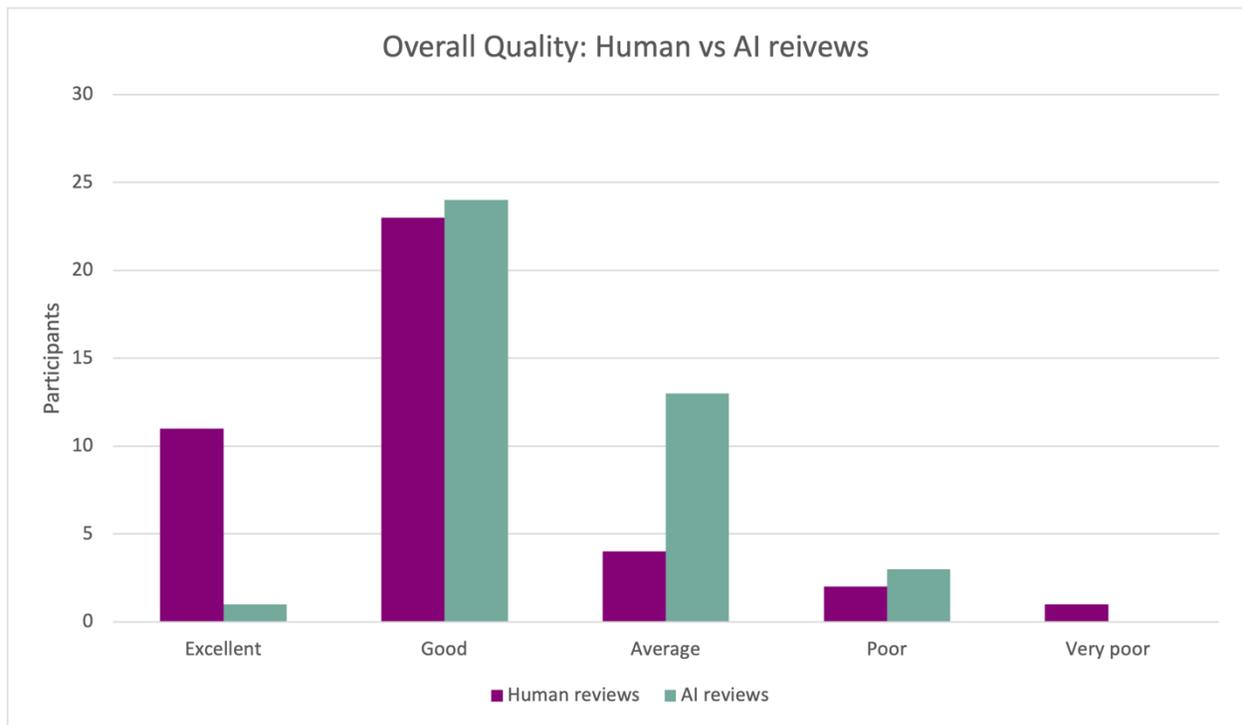

*Fig. 16: Overall quality: Human vs AI reviews*

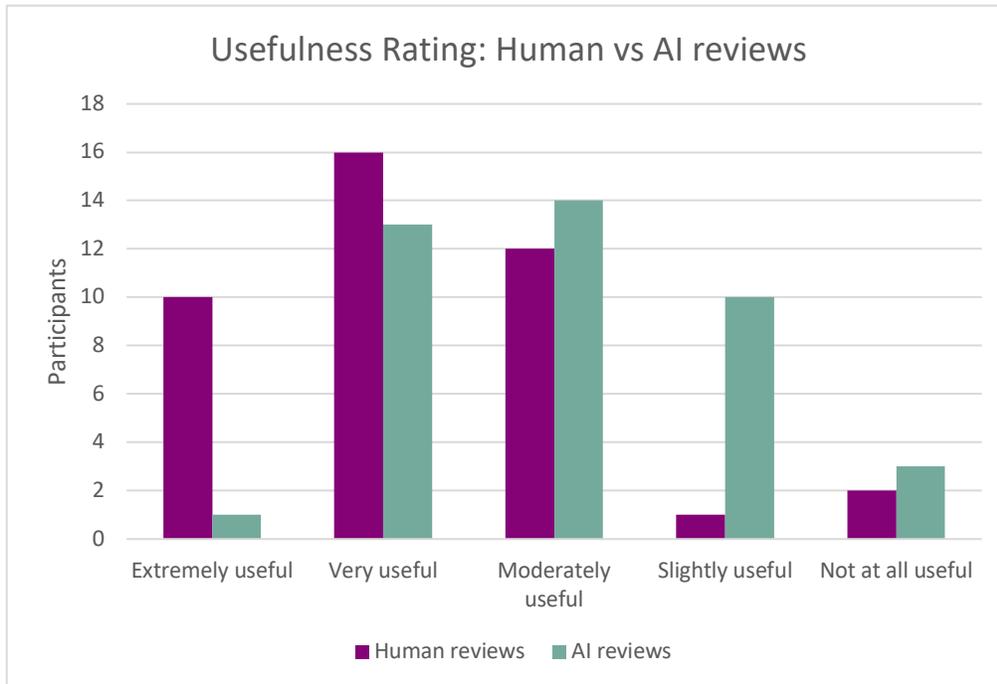

*Fig. 17: Usefulness Rating: Human vs AI reviews*

When asked to share the advantages of the AI review, the majority of the participants (n=31) said that the AI feedback provided instant and consistent reviews aligned with the rubric: *"Quick feedback, rubric specific feedback"* (Participant 30). Some participants (n=11) considered useful that the AI feedback was rubric-based, as it helped them to understand how their work related to the rubric requirements. Participant 43 noted that *"you can review the rubrics and criteria"* and *"generally know what to revise to meet the requirement."* Participants also appreciated the impartiality of the AI's evaluation (n=20), as well as its ability to catch macro issues and provide thorough summaries with specific and clear comments, as participant 13 stated: *"the details provided by the AI review were very concise yet detailed."*

Regarding the disadvantages that participants identified about the AI reviews, most participants (n=27) expressed their concerns about the feedback being too general, lacking the human touch that the peer reviewers could provide, leading to difficulties in identifying specific areas for improvement. For instance, participant 36 pointed out that *"AI systems may not be as effective as people in evaluating papers in certain specialized domains...People can identify subtle flaws and provide feedback on how to improve the paper/work project."* Few participants also expressed concerns about the accuracy of the AI review (n=7), especially when the AI failed to identify important details in the work. They noted that the AI review *"missed some smaller points"* (Participant 34) and *"was not specific enough in terms of items that needed to be revised"* (Participant 38).

Moving forward to the benefits of human reviews identified in the survey, 19 participants mentioned the importance of receiving specific and constructive feedback from their peers. Participant 14 noted that peer reviewers provided *"on point and specific"* feedback that suggested insightful recommendations. Many participants (n=26) also appreciated the fresh perspective that peer reviews provided, with participant 28 noting that peer review *"reveal[s] blind spots"* and offers *"diverse feedback."* Most participants (n=32) valued the depth and nuance offered by human reviewers, with participant 36 mentioning that peer reviewers *"can identify subtle flaws and provide feedback on how to improve the paper/work project"*.

Concerning the disadvantages of human reviews through CGMap, the most significant insight is that human reviewers are considered subjective and can be inconsistent, influenced by their personal biases and preferences (n=15). For instance, Participant 36 stated that *"Human reviewers may have different levels of experience, expertise, and biases, which can lead to inconsistent evaluations of papers/work projects."* Another prevalent insight is that there can be a delay in receiving feedback, and sometimes, reviews may not be received at all (n=26), as participant 30 said, *"not everyone reviewed, time."*

The survey data show that the participants' overall preference lean towards the peer reviews (n=18) or a combination of both peer and AI reviews (n=19). Participants appreciate the specific and actionable feedback provided by their peers, as well as the opportunity to collaborate and get customized comments, as participant 37 mentioned *"I think the human reviews are written by people who have shared the course experience and understand the demands and limitations of these works."* Participants recognize that AI reviews have the potential to provide useful feedback, but there are still limitations in making it nuanced, accurate, and personalized. That is why a combination of both review types is considered the most effective, as the two review types overlap by omitting each other disadvantages, as participant 19 highlighted, *"the combination of both the human and AI work is helpful in having both general and nuanced feedback for my work"*.

## 4. Discussion

"Large Language Models," so-called, are still language models in Church and Mercer's sense (Church and Mercer 1993)—they have no understanding of language. Machine intelligence records characters in binary notation, and to perform statistical analyses of patterns in these characters. Limited though this is as mechanized intellectual work, the sheer force of numerical modelling and calculation can today achieve feats that are superhuman. In addition to responding to prompts, they perform second order operations including summarization, translation, application of instructions, and planning. The capabilities of generative AI are "emergent" and not even fully explicable even to the engineers who built them.

The power of this unmeaning brute force was clearly evident to the participants in our reported in this paper. The AI reviews come back to students as neatly framed narrative responses. Participants reported that AI reviews were helpfully different from human feedback in a number of ways, even against identical prompts.

One key to the implementation of C-LLMs, we propose, lies in a new domain of human-computer interaction termed "prompt engineering." We outlined in the second section of this paper the serious and intrinsic deficiencies of C-LLMs in the areas of sourcing, facts, theory, ethics and critical dialogue—crucial epistemic virtues and pedagogical groundings. These deficiencies have been confirmed by the course participants in this intervention, themselves experienced educators undertaking graduate programs.

The main strength of C-LLMs is that they are good at spinning into narrative form texts that draw from only sometimes but not necessarily from reliable sources, using possible facts, applying possible theories, and viewing these through the lens of possibly critical analysis. The machine has no way of knowing whether its sources are reliably sourced, factual, theoretically sound or survive the rigors of critical interrogation. What we need for reliable knowledge work and good learning is to feed the machine with the epistemic virtues of using reliable sources and resilient facts, theories, and critical perspectives.

We do this in CGMap in two ways. First, the generative AI is presented student texts that have already been vetted by peers for these epistemic virtues. Then, second, we use generative AI to provide supplementary narrative reviews through careful prompt engineering. In a sense, we have told the GPT in general terms what to say in response to the specifics of the student texts.

In CGMap, we've set out to develop software that recalibrates C-LLMs to make them as useful as possible for learning. We present these recalibrations as three frames:

1. *An Epistemic Frame: prompt the machine to offer students feedback on the basis of a theory of knowledge applicable to their learning.* In our experiment, we used narrative elaborations via a rubric framed in terms of the "knowledge processes" of our epistemological theory of learning. CGMap then runs through each piece of student work multiple times, offering narrative feedback framed by epistemological criteria embedded in the rubric.

2. *An Empirical Frame: require the learners to bring verifiable facts to the machine.* We prompt the C-LLM to respond with anything factual because it is a "black box" that fails to acknowledge its sources and cannot know fact from fake. The narrative generated by the prompt is only valid to the extent that it works with the facts that it has been fed, already verified in human peer reviews.

3. *An Ontological Frame: bring the theoretical frames of disciplines to the machine.* In an extension of our recent work in the area of medical education (Cope et al. 2022), we are applying the formal ontologies of biomedical practice to the prompt, not as circumstantially collocated clusters of characters, but the widely agreed definitions and taxonomically well-formed schemas that define the medical domain. Many academic fields are supported by such schemas, frequently systematized in XML markup in addition to the metadata schemas that drive everyday interoperability across computer applications and the internet (Cope and Kalantzis 2023a). Curriculum standards documents are in computational terms discipline-specific ontologies. CGMap has been designed to ingest any well-formed ontological frame.

To conclude now with two propositions, in tension. On the one hand, the generative AI of C-LLMs is architected in a way that is worse-than harmful to education. It undermines some of the key epistemological bases of modern science and reliable knowledge systems. (A separate question is, did the technology have to turn out this way? Our tentative answer is, perhaps not if they were architected along lines of the recalibrations we have created in CGMap.)

On the other hand, the allure of C-LLMs is their neatly formed, speedy narrative responses. These characteristics were particularly highlighted by our participants. With epistemic, empirical and ontology-based recalibration, C-LLMs can offer feedback to learners that usefully supplements human feedback. Besides, C-LLMs have "read" massive bodies of digitized text, of considerable value in itself, even if their abilities to distinguish reliable fact and theory and to determine the credibility of sources are non-existent, and their ethical and critical faculties are at best questionable. At least they are interesting interlocutors, thought-provoking even for their untrustworthiness.

So what can C-LLMs be used for? Our answer is: much less than they implicitly purport to do when they respond to a prompt. But now that it's here, generative AI is not going to go away. Attempts to ban it or slow its development are doomed. Purposefully recalibrated, we contend, these stochastic parrots can be put to good use supporting learning, so long as their role is confined to what have discursive narrative—tying independently verified credible facts, theories and sources and into well-framed discourse.

Like all parrots, what C-LLMs say is only as good as what we tell them to say. To tell generative AI what to say, we educators must now become prompt engineers. And of course, any willing and agreeable interlocutor soon becomes likable. If it can help learning, we may come to like this particular parrot quite a lot.

Laurillard, Diana, *Teaching as a Design Science: Building Pedagogical Patterns for Learning and Technology*, London: Routledge, 2012.

Li, Zhixin and Ying Xu, "Designing a Realistic Peer-like Embodied Conversational Agent for Supporting Children's Storytelling," *arXiv,* 2304.09399, 2023, doi: https://doi.org/10.48550/arXiv.2304.09399.

Lim, Fei Victor, Bill Cope and Mary Kalantzis, "A Metalanguage for Learning: Rebalancing the Cognitive with the Socio-Material," *Frontiers in Communication,* 7(Article 830613):1-15, 2022, doi: http://doi.org/10.3389/fcomm.2022.830613.

Liu, Yikang, Ziyin Zhang, Wanyang Zhang, Shisen Yue, Xiaojing Zhao, Xinyuan Cheng and Hai Hu Yiwen Zhang, "ArguGPT: Evaluating, Understanding and Identifying Argumentative Essays Generated by GPT Models," *arXiv,* 2304.07666 2023, doi: https://doi.org/10.48550/arXiv.2304.07666.

Luckin, Rosemary, *Machine Learning and Human Intelligence: The Future of Education in the 21st Century*: UCL Institute of Education Press, 2018.

Luo, Junwen, Thomas Feliciani, Martin Reinhart, Judith Hartstein, Vineeth Das, Olalere Alabi and Kalpana Shankar, "Analyzing Sentiments in Peer Review Reports: Evidence from Two Science Funding Agencies," *Quantitative Science Studies,* 2(4):1271–95, 2022, doi: https://doi.org/10.1162/qss_a_00156.

Magee, Liam, Vanicka Arora and Luke Munn, "Structured Like a Language Model: Analysing AI as an Automated Subject," *arXiv,* 2212.05058, 2022, doi: https://doi.org/10.48550/arXiv.2212.05058.

Markauskaite, Lina, Rebecca Marrone, Oleksandra Poquet, Simon Knight, Roberto Martinez-Maldonado, Sarah Howard, Jo Tondeur, Maarten De Laat, Simon Buckingham Shum, Dragan Gašević and George Siemens, "Rethinking the Entwinement Between Artificial Intelligence and Human Learning: What Capabilities Do Learners Need for a World with AI?," *Computers and Education: Artificial Intelligence,* 3:1-16, 2022, doi: https://doi.org/10.1016/j.caeai.2022.100056.

Markel, Julia M., Steven G. Opferman, James A. Landay and Chris Piech, "GPTeach: Interactive TA Training with GPT Based Students," *EdArXiv*, 2023, doi: https://doi.org/10.35542/osf.io/r23bu.

McCarthy, John, Marvin L. Minsky, Nathaniel Rochester and Claude E. Shannon, *A Proposal for the Dartmouth Summer Research Project on Artificial Intelligence*, 1955.

Mori, Masahiro, "The Uncanny Valley," *IEEE Robotics & Automation Magazine,* 19(2):98-100, 1970 [2012], doi: https://doi.org/10.1109/MRA.2012.2192811.

Munn, Luke, Liam Magee and Vanicka Arora, "Truth Machines: Synthesizing Veracity in AI Language Models," *arXiv,* 2301.12066, 2023, doi: https://doi.org/10.48550/arXiv.2301.12066.

New London Group, "A Pedagogy of Multiliteracies: Designing Social Futures," *Harvard Educational Review,* 66(1):60-92, 1996, doi: https://doi.org/10.17763/haer.66.1.17370n67v22j160u.

Newman, Heather and David Joyner, "Sentiment Analysis of Student Evaluations of Teaching," paper presented at the International Conference on Artificial Intelligence in Education, Cham CH, 2018, doi: https://doi.org/10.1007/978-3-319-93846-2_45.

Nilsson, Nils J., *The Quest for Artificial Intelligence: A History of Ideas and Achievements*, Cambridge UK: Cambridge University Press, 2009.
33